\documentclass[usenatbib]{mnras}
\usepackage{graphicx}
\usepackage{natbib}
\usepackage{latexsym}
\usepackage{amsmath}
\usepackage{setspace}
\usepackage{booktabs}
\usepackage{array, longtable}
\usepackage[bf, singlelinecheck=off]{caption}
\usepackage{subfig}
\usepackage[T1]{fontenc}
\usepackage{aecompl}
\usepackage{floatrow}
\newcommand{\etal}{et~al.}

\newcommand{\kms}{$\mbox{km~s}^{-1}$}

\newcommand{\specsfig}[1]        
{
   \begin{center}
     \begin{minipage}[t]{0.45\textwidth}
         \psfig{file=#1.eps,height=0.9\textwidth,angle=270}
     \end{minipage}
     \end{center}
 }

\newcommand{\specdfig}[2]        
{
   \begin{center}
     \begin{minipage}[t]{0.45\textwidth}
         \psfig{file=#1.eps,height=0.9\textwidth,angle=270}
     \end{minipage}
     \hfill
     \begin{minipage}[t]{0.45\textwidth}
         \psfig{file=#2.eps,height=0.9\textwidth,angle=270}
     \end{minipage}
   \end{center}
}

\def \la{\mathrel{\mathchoice   {\vcenter{\offinterlineskip\halign{\hfil
					$\displaystyle##$\hfil\cr<\cr\sim\cr}}}
		{\vcenter{\offinterlineskip\halign{\hfil$\textstyle##$\hfil\cr
					<\cr\sim\cr}}}
		{\vcenter{\offinterlineskip\halign{\hfil$\scriptstyle##$\hfil\cr
					<\cr\sim\cr}}}
		{\vcenter{\offinterlineskip\halign{\hfil$\scriptscriptstyle##$\hfil\cr
					<\cr\sim\cr}}}}}
\def \ga{\mathrel{\mathchoice   {\vcenter{\offinterlineskip\halign{\hfil
					$\displaystyle##$\hfil\cr>\cr\sim\cr}}}
		{\vcenter{\offinterlineskip\halign{\hfil$\textstyle##$\hfil\cr
					>\cr\sim\cr}}}
		{\vcenter{\offinterlineskip\halign{\hfil$\scriptstyle##$\hfil\cr
					>\cr\sim\cr}}}
		{\vcenter{\offinterlineskip\halign{\hfil$\scriptscriptstyle##$\hfil\cr
					>\cr\sim\cr}}}}}

\begin{document}
\title[Variability in extragalactic class I methanol masers]{Variability in extragalactic class I methanol masers: New maser components toward NGC~4945 and NGC~253}
\author[McCarthy \etal]{
	T.\ P. McCarthy,$^{1,2}$\thanks{Email: tiegem@utas.edu.au}
	S.\ P. Ellingsen,$^{1}$
    S. L. Breen,$^{3}$ 
	M.\ A. Voronkov,$^{2}$ 
	X. Chen$^{4,5}$ and
	\newauthor H-h. Qiao$^{5,6}$ \\
  \\
  $^1$ School of Natural Sciences, University of Tasmania, Private Bag 37, Hobart, Tasmania 7001, Australia\\
  $^2$ Australia Telescope National Facility, CSIRO, PO Box 76, Epping, NSW 1710, Australia \\
  $^3$ Sydney Institute for Astronomy (SIfA), School of Physics, University of Sydney, NSW 2006, Australia\\
  $^4$ Center for Astrophysics, GuangZhou University, Guangzhou 510006, China\\
  $^5$ Shanghai Astronomical Observatory, Chinese Academy of Sciences, Shanghai 200030, China \\
  $^6$ National Time Service Center, Chinese Academy of Sciences, Xi'An, Shaanxi 710600, China \\}

 \maketitle

\begin{abstract}
We have used the Australia Telescope Compact Array (ATCA) to make new observations of the 36.2-GHz ($4_{-1}\rightarrow3_0$E) methanol transition toward NGC~4945 and NGC~253. These observations have revealed the presence of new maser components toward these galaxies, and have provided the first clear evidence for variability in extragalactic class~I methanol masers. Alongside the new observations of NGC~4945 and NGC~253, we present the results of recent 36.2-GHz methanol maser searches toward 12 galaxies, placing upper limits on the emission from the 36.2-GHz class~I transition and the 37.7-GHz  ($7_{2}\rightarrow8_1$E) class~II maser line toward these sources. Flux density values for the 7-mm continuum emission toward these sources are also reported where applicable. A re-analysis of the published 36.2-GHz methanol observations of Arp 220 undertaken as part of the search revealed some issues with previous imaging procedures.  The re-analysis, combined with non-detections in independent follow-up observations suggest that there is no 36.2-GHz methanol emission toward Arp~220 stronger than 3.5 mJy in a 10~\kms\ channel (5$\sigma$ upper limit).
\end{abstract}


\section{Introduction}

The properties of galaxies (particularly those that are very distant) are often unable to be directly measured. Instead, we rely on indirect measurements through observations of some proxy. Historically, maser emission has proven to be a powerful tool for both direct and indirect investigation of various astrophysical phenomena both within our Galaxy and in others \citep[e.g. star-formation, outflows, black-hole accretion disks;][]{Moran+99, Kurtz+04}. Galactic methanol masers, in particular, allow insight into the various stages of star-formation regions. Their compact and generally highly luminous nature, allows for conditions or evolutionary stages to be detected and accurately ascribed to certain locations within these star-forming regions \citep[e.g.][]{Cyganowski+09,Cyganowski+12,Breen+13b,Voronkov+14}.  Extragalactic methanol masers may prove to be a similarly interesting tool for investigation of external galaxies, however, at this stage they are a relatively newly discovered and not well understood phenomenon.  

The majority of known extragalactic methanol masers have been detected in the past five years. Class~I methanol masers have been reported in NGC 253, Arp~220, NGC 4945, NGC 1068, IC~342 and NGC~6946 \citep{Ellingsen+14, Wang+14, Chen+15, McCarthy+17, McCarthy+18b, Gorski+18} and class~II masers in the Large Magellanic Cloud (LMC) and M31 \citep{Green+08,Ellingsen+10,Sjouwerman+10}. Considering these extragalactic examples, the class~II masers appear to be highly luminous examples of regular Galactic-type class~II masers \citep{Ellingsen+10}, however, this is not the case with the current detections of class~I masers. Instead, they appear to be associated with large-scale regions of low-velocity shocks, such as molecular in-fall, within their host galaxies \citep{Ellingsen+17b,Gorski+18,McCarthy+18c}.

NGC~253 and NGC~4945 are two nearby, southern, barred spiral galaxies with starburst nuclei. These sources are considered the prototypical examples of extragalactic class~I methanol masers. The methanol masers in these sources are associated with the galactic bars of their host. The masers in NGC~253 appear to be associated with the inner bar-nucleus interface, where molecular in-fall along the bar is interacting with the central starburst zone, resulting in large-scale low velocity shocks \citep{Ellingsen+17b}. The single maser region in NGC~4945 appears to be located at the interface region between the galactic bar and the south-eastern spiral arm \citep{Henkel+18, McCarthy+18c}.

Currently, there is no direct evidence for variability toward any extragalactic methanol maser sources. In Galactic sources, variability has been detected in numerous class~II masers \citep[e.g.][]{Goedhart+03, Olech+19}. However, the same has not been reported for Galactic class~I sources. Variability in these class~II sources has been suggested to be related to either changes in the pump rate \citep{Araya+10}, or variations in seed photon flux of the masers \citep{vanderwalt11}.

In this paper we report the results of new 7-mm observations toward NGC~4945 and NGC~253. Additionally we report upper-limits on the luminosity of 36.2-GHz class~I and 37.7-GHz class~II methanol emission toward twelve extragalactic sources (see Table \ref{tab:source_details}), from data taken with the ATCA over the past few years. The initial search for extragalactic class~I methanol masers was made towards a sample of luminous extragalactic OH and H$_2$O megamasers (9 of the 12 sources reported here). These targets were selected before the first detection of an extragalactic class~I methanol maser had been reported. This initial search was responsible for the first successful detections of 36.2-GHz class~I masers reported in \citet{Ellingsen+14, Ellingsen+17b} and \citet{McCarthy+17}. The remaining three sources (NGC~1097, NGC~1792 and NGC~1808) were observed as part of a follow up search, and as such selected targets are based on the general properties of NGC~253 and NGC~4945 (elevated star-formation rates, barred spirals etc.). We also provide a discussion of the entire sample, both detections and non-detections, which will allow a more focussed approach to target selection, allowing a higher success rate for future searches.


\begin{table*} 
	\begin{center}
		\caption{Details for target sources. Recession velocities have been taken from the \textit{NASA/IPAC Extragalactic Database}$^\dagger$~(NED). Redshift-independent distances have been provided where possible, otherwise distances reported are the luminosity distances provided by NED. }
		\begin{tabular}{@{}lcclll@{}}
			\toprule
			\multicolumn{1}{c}{Target}  & \multicolumn{1}{c}{Right Ascension}  &\multicolumn{1}{c}{Declination}     & \multicolumn{1}{c}{$V_{helio}$} & \multicolumn{1}{c}{$D$} & \multicolumn{1}{c}{Type} \\
			\multicolumn{1}{c}{Source} &   \multicolumn{1}{c}{\bf $h$~~~$m$~~~$s$}& \multicolumn{1}{c}{\bf $^\circ$~~~$\prime$~~~$\prime\prime$} &\multicolumn{1}{c}{(\kms)}  &\multicolumn{1}{c}{(Mpc)} &  \multicolumn{1}{c}{}  \\ \midrule
			NGC\,253             & 00 47 33.10 & --25 17 18.00 & 243 & 3.4$^{[1]}$  & SAB(s)c \\
			IRAS\,01417+1651         & 01 44 30.50 & +17 06 05.00 & 8214 & 119 & pair of galaxies \\
			NGC\,1097             & 02 46 19.06  & --30 16 29.68 & 1271 & 9.51$^{[2]}$  & SB(s)b, Sy1 \\
			NGC\,1792             & 05 05 14.45 & --37 58 50.70 & 1211 & 9.6$^{[3]}$   & SA(rs)bc\\
			NGC\,1808             & 05 07 42.34 & --37 30 46.98 & 995 & 6.8$^{[3]}$  & (R$^\prime$)SAB(s)a, Sy2 \\
			ESO\,558-g009           & 07 04 21.02 & --21 35 19.20 & 7674 & 118 & Sc \\
			IC\,2560                 & 10 16 18.70 & --33 33 50.00 & 2925 & 32.5$^{[4]}$ & (R$^\prime$)SB(r)b \\
			IRAS\,10173+0829      & 10 20 00.20 & +08 13 34.00 & 14716 & 231 & LIRG \\
			IRAS\,11506-3851         & 11 53 11.70 & --39 07 49.00 & 3232 & 30.3$^{[5]}$ & (R$^\prime$)SAB(r)a \\
			ESO\,269-g012             & 12 56 40.50 & --29 34 02.81 & 5014 & 78.8 & S0, Sy2 \\
			NGC\,4945             & 13 05 27.50 & --49 28 06.00 & 563 & 3.7$^{[6]}$  & SB(s)cd, Sy2 \\
			Circinus               & 14 13 09.30 & --65 20 21.00 & 434 & 4.21$^{[7]}$ & SA(s)b \\
			NGC\,5793                & 14 59 24.70 & --16 41 36.00 & 3491 & 55 & Sb, Sy2 \\
			IIZw\,96                     & 20 57 23.90 & +17 07 39.00 & 10822 & 159  & pair of galaxies \\ \hline                  
		\end{tabular} \label{tab:source_details}
	\end{center}
\begin{flushleft}
	Note: $^{[1]}$\citet{Dalcanton+09}, $^{[2]}$\citet{Bottinelli+85}, $^{[3]}$\citet{Sorce+14}, $^{[4]}$\citet{Yamauchi+12}, $^{[5]}$\citet{Tully+16}, $^{[6]}$\citet{Tully+13} and $^{[7]}$\citet{Tully+09} \\
\end{flushleft}	
\end{table*}

\begin{table*} 
	\begin{center}
		\caption{Details of the observations for all target sources reported in this paper. The asterisk denotes an observation epoch previously reported in \citet{McCarthy+18c}. The details for this epoch have been reproduced here as it is important for the discussion of variability in Section \ref{sec:variability}.}
		\begin{tabular}{@{}llllllc@{}}
			\toprule
			\multicolumn{1}{c}{Target}  & \multicolumn{1}{c}{Epoch}  &\multicolumn{1}{c}{Array}     & \multicolumn{1}{c}{Flux} & \multicolumn{1}{c}{Bandpass} & \multicolumn{1}{c}{Phase} &\multicolumn{1}{c}{On-source} \\
			\multicolumn{1}{c}{Source} &             & \multicolumn{1}{c}{Configuration} &\multicolumn{1}{c}{Calibrator}       & \multicolumn{1}{c}{Calibrator} & \multicolumn{1}{c}{Calibrator}    & \multicolumn{1}{c}{Time}     \\ \midrule
			NGC\,253             & 2019 March & H214  &  PKS\,B1934-638   & PKS\,B1921-293   & 0116-219   & 4.84h \\
			IRAS\,01417+1651                & 2014 March & H168  &  Uranus & PKS\,B1921-293 & 0221+067 & 0.50\,h \\
			NGC\,1097              & 2018 November & H168  &  PKS\,B1934-638   & PKS\,B1921-293   &  0237-233  & 4.61h \\
			NGC\,1792             & 2018 December & H168  &  PKS\,B1934-638   & PKS\,B1921-293   & 0521-365   &  3.72h \\
			NGC\,1808              & 2018 November & H168  &  PKS\,B1934-638   & PKS\,B1921-293   &  0521-365  & 1.97h \\
			ESO\,558-g009                 & 2014 March  & H168   & Uranus  & PKS\,B1921-293 & 0648-165 & 0.49\,h \\
			IC\,2560                      & 2014 March  &  H168   & PKS\,B1934-638  &  PKS\,B1253-055  & 1034-293  &  0.20\,h  \\
			IRAS\,10173+0829              & 2015 August & EW352  &  PKS\,B1934-638   & PKS\,B1921-293   &  1036+054  & 4.02h \\
			IRAS\,11506-3851                & 2014 March  &  H168   & PKS\,B1934-638 & PKS\,B1253-055 & 1144-379 & 0.50\,h \\
			ESO\,269-g012                 & 2014 March  &  H168    & PKS\,B1934-638 & PKS\,B1253-055  & 1322-427  &    0.20\,h                \\
			NGC\,4945  	& 2017 June* & H214  &  PKS\,B1934-638   & PKS\,B1921-293   & j1326-5256   & 1.64h \\           
			& 2017 July & H75  &  Uranus   & PKS\,B1253-055   & j1326-5256   & 1.92h \\
			& 2019 March & H214  &  PKS\,B1934-638   & PKS\,B1921-293   & j1326-5256   & 3.93h \\ 
			Circinus                    & 2015 August &  EW352  & PKS\,B1934-638    &  PKS\,B1921-293   & 1414-59 &  2.80\,h \\
			NGC\,5793                     & 2014 March  &  H168  & PKS\,B1934-638 & PKS\,B1253-055  & 1510-089 & 0.20\,h \\
			IIZw\,96                      & 2015 August & EW352  & PKS\,B1934-638  &  PKS\,B1921-293     &   2121+053     & 2.32\,h \\ \hline                  
		\end{tabular} \label{tab:observations}
	\end{center}
\end{table*} 

\section{Observations} \label{sec:observations}

The 7-mm observations reported in this paper were all made using the Australia Telescope Compact Array (ATCA) across multiple epochs between 2014 and 2019 (see Table \ref{tab:observations}). We adopted rest frequencies of 36\,169.238$\pm 0.001$\,MHz \citep{Voronkov+14} and 37\,703.700$\pm$0.030\,MHz \citep{Tsunekawa+95} for the class~I 36.2- and class~II 37.7-GHz transitions of methanol respectively. 

Our observations are single pointings toward the centre of each galaxy (see Table \ref{tab:source_details}). At 36.2-GHz the FWHM of the ATCA primary beam is 84 arcseconds. This corresponds to a linear radius from the pointing centre of $\sim$700~pc for our nearest source (NGC~253 at 3.4~Mpc) and $\sim$47~kpc for our furthest (IRAS\,10173+0829 at 231~Mpc). This range is sufficient for detecting class~I methanol masers at similar spatial offsets from the nucleus of the galaxies as those reported in previous reported detections.

Either PKS\,B1921--293 or PKS\,B1253--055 was utilised as a bandpass calibrator dependent on the epoch. The 10 minute scans of the target source were interleaved with 2 minute phase calibrator scans throughout the observations, with flux density calibration using PKS\,B1934--638 and Uranus. All sources observed in each epoch, and their relevant calibrators are listed in Table \ref{tab:observations}. 

The details of the Compact Array Broadband Backend \citep[CABB;][]{Wilson+11} configurations for each epoch are detailed in the following subsections.

\subsection{2014 March} \label{sec:march_deets}

Observations were made on 2014 March 26 and 28 (project code C2879), using the H168 hybrid array configuration (minimum and maximum baselines of 61 and 192m respectively). CFB 1M CABB mode was utilised for these observations, resulting in $2 \times 2048$ MHz IF bands, with centre frequencies of 35.3\,GHz and 37.3\,GHz respectively. The channel width for these IF bands was 1\,MHz ($\sim$8.2~\kms\ at 36.2-GHz) resulting in 2048 spectral channels for each IF band. A typical synthesised beam for this hybrid array at 36.2-GHz is approximately $9''\times5''$ (dependent on source declination).

Dependent on the day, the observing strategy utilised PKS\,B1934-638 or Uranus as flux density calibrators, with bandpass calibrated with respect to PKS\,B1921-293 or PKS\,B1253-055 (see Table \ref{tab:observations}). Scans of target sources were interleaved with phase calibration scans on nearby appropriate sources. 

The weather for this epoch of observation was particularly bad which, when combined with low on-source times, resulted in bad data quality. We have only reported detection upper-limits for those sources which we could image to a sufficient standard. This generally includes sources where we could get continuum detections (see Table \ref{tab:continuum}) in order to self-calibrate the data and achieve higher sensitivity. As an example of the data quality, NGC~4945 was originally observed during this epoch, and the data quality is too poor to detect the strong 36.2-GHz emission observed toward this source in observations since. A table of sources that were also observed, but with too low data quality (RMS over an order of magnitude higher than expected) to draw any useful conclusions, is included in the appendix (Table \ref{tab:excluded_sources}).

\subsection{2015 August}

Observations were made on 2015 August 25 and 26 (project code C2879), using an EW352 east-west array configuration (minimum and maximum baselines of 31 and 352m respectively). The CABB was configured in the CFB 1M/64M hybrid mode, consisting of $2 \times 2048$ MHz IF bands. The first IF band has a channel width of 1 MHz resulting in 2048 spectral channels, with a centre frequency for the band of 36.85 GHz. The second IF is $32\times64$\,MHz channels (also centred on 36.85 GHz) with the ability to define up to 16$\times$2048 channel zoom bands with a fine resolution of 32\,kHz (resulting in a channel width of $\sim$0.26~\kms\ at 36.2-GHz). Two sets of five zoom bands were combined (`stitched') together to cover each of the two lines of primary interest (36.2-GHz class~I and 37.7-GHz class~II methanol). A typical synthesised beam for these 36-GHz EW352 observations is $26'' \times 4''$.


\subsection{2017 July}

Observations were made on 2017 July 7 (project code C3617), using the ATCA H75 hybrid array configuration (minimum and maximum baselines of 30.6 and 89.2m respectively). The CABB was configured in CFB 1M-0.5k mode, resulting in $2 \times 2048$ MHz IF bands (with centre frequencies of 36.85 and 32.20 GHz), each with 1 MHz spectral resolution ($\sim$8.2~\kms\ at 36.2-GHz) and the option of configuring 16 zoom bands per IF, with 2048 channels and a fine resolution of 0.5~kHz ($<0.01$~\kms\ at 36.2-GHz). We did not use zoom bands for these observations, due to the fine resolution producing insufficient velocity coverage for extragalactic sources at this frequency. These H75 array observations at 36.2-GHz had a synthesised beam size of approximately $27'' \times 12''$.


\subsection{2018 November and December} \label{sec:2018_nov_dec}

Observations were made on 2018 November 28 and 29, and December 3 (project code C3263) using the H168 hybrid array configuration (minimum and maximum baselines of 61 and 192m respectively). The CFB 64M-32k CABB configuration was used, consisting of two $32\times64$\,MHz IF bands (centre frequencies of 32.2 and 36.85 GHz respectively), with both bands additionally allowing up to 16$\times$2048 channel zoom bands to be defined with a fine resolution of 32\,kHz. Each target methanol line (36.2-GHz class~I and 37.7-GHz class~II methanol), was covered by 4 stitched zoom bands (the 32 kHz channel width corresponds to $\sim$0.26 \kms\ at this frequency). The synthesised beams of these observations are approximately $6''\times4''$.


It should be noted that due to frequency configuration issues, the flux density calibration for NGC~1097 required approximation based on levels seen from the other recent observations. NGC~1097 and NGC~1808 were observed one day apart and comparison of the flux density values for the bandpass calibrator used for both these sets of observations (PKS\,B1921--293) allowed for an approximate flux density calibration for NGC~1097.  Therefore, both RMS values reported in Table \ref{tab:cubes} and continuum flux density values in Table \ref{tab:continuum} have higher levels of uncertainty ($\sim$40 percent) than all other sources.

\subsection{2019 March}

Observations were made on 2019 March 3 (project code C3167), using the H214 hybrid array configuration minimum and maximum baselines of 82 and 247 m respectively). The CFB 64M-32k CABB configuration was used, consisting of two $32\times64$\,MHz IF bands (centre frequencies of 36.8 and 38.00 GHz respectively), with both bands additionally allowing up to sixteen 2048 channel zoom bands to be defined with a fine resolution of 32\,kHz. The target 36.2-GHz methanol line was covered by 4 stitched zoom bands (the 32 kHz channel width corresponds to $0.26$ \kms\ at this frequency). These observations had a synthesised beam of $6\farcs3\times4\farcs4$ and $5\farcs6\times4\farcs3$ for NGC~4945 and NGC~253 respectively.

The bandpass calibrator PKS\,B1921-293 was used with flux density calibrated against PKS\,B1934-638. A nearby quasar was selected for interleaved phase calibration scans for each source. Details on these sources can be found in Table \ref{tab:observations}.

The weather during this session was not ideal for mm-wavelength observing, with high ambient temperatures and patchy cloud cover. This results in higher RMS noise levels than would be expected for our on-source time. However, a combination of manual flagging (for scans affected by clouds passing through the line of sight) and self-calibration allows us to achieve a sensitivity similar to past observations of these sources \citep{McCarthy+18c, Ellingsen+17b}.

\begin{table} 
	\begin{center}
		\caption{Details for the spectral line cubes used to search for methanol spectral lines. Sources highlighted in bold text are those from project C3167, while the rest are from the maser search projects. Velocity ranges and channel widths are identical for the 36.2- and 37.7-GHz spectral lines. The two RMS noise values correspond to the noise for the 36.2- and 37.7-GHz cubes respectively (not applicable to C3167 sources where only 36.2-GHz is observed). RMS noise values are reported post self-calibration where relevant. The RMS values for NGC~1097 (denoted by an asterisk) have a higher level of uncertainty due to less accurate flux density calibration (see Section \ref{sec:2018_nov_dec}). }
		\begin{tabular}{@{}lcll@{}}
			\toprule
			\multicolumn{1}{c}{Target}  & \multicolumn{1}{c}{Velocity}  &\multicolumn{1}{c}{Channel}     & \multicolumn{1}{c}{RMS}  \\
			\multicolumn{1}{c}{Source} &   \multicolumn{1}{c}{Range}   & \multicolumn{1}{c}{Width} &\multicolumn{1}{c}{Noise}      \\
			\multicolumn{1}{c}{} &   \multicolumn{1}{c}{(\kms)}   & \multicolumn{1}{c}{(\kms)} &\multicolumn{1}{c}{(mJy beam$^{-1}$)}      \\ \midrule
			\textbf{NGC\,253}              & 0--500  &  1 &  1.9 \\ 
			IRAS\,01417+1651                &  7100 -- 9100 & 10 &  1.6/1.6 \\
			NGC\,1097             & 750 -- 1750 & 10  &  0.5*/0.4*   \\
			NGC\,1792             &  700 -- 1700 & 10  &  0.4/0.4  \\
			NGC\,1808              & 600--1600  &  10 &  0.2/0.2 \\
			ESO\,558-g009           &  6800 - 9600 & 10 & 2.2/2.2 \\
			IC\,2560                 & 2000 -- 3600  &  10 & 6.5/6.7 \\
			IRAS\,10173+0829       & 13500 -- 14900  &  10  &  1.2/1.3   \\
			IRAS\,11506-3851         & 1900 -- 4100  &  10 &  2.2/2.5 \\
			ESO\,269-g012             &  4000 -- 6000 & 10  & 6.6/6.2  \\
			\textbf{NGC\,4945 (H214)}              & 250--850  &  1 &  1.9 \\
			\textbf{NGC\,4945 (H75)}              &  200--1020 &  8.2 & 0.3 \\
			Circinus                & 0 -- 800 &  10 &  0.5/0.6 \\
			NGC\,5793                &  2500 -- 4400 & 10 & 6.4/5.4   \\
			IIzw96                     & 10000 -- 12000  & 10 &  0.6/0.6   \\ \hline           
		\end{tabular} \label{tab:cubes}
	\end{center}
\end{table} 

\begin{table*} 
	\begin{center}
		\caption{NGC~4945 36.2-GHz methanol emission details from the 2019 March H214 observations. All reported velocities are with respect to the barycentric reference frame. Uncertainties in the reported flux density values result from the random errors in fitting using the {\sc miriad} task imfit. Please note the integrated flux density for the continuum emission has units of mJy rather than mJy~\kms. }
		\begin{tabular}{@{}cllllccc@{}}
			\toprule
			 \multicolumn{1}{c}{ \bf$\#$} & \multicolumn{1}{c}{Label}  &\multicolumn{1}{c}{RA (J2000)}     & \multicolumn{1}{c}{Dec (J2000)} & \multicolumn{1}{c}{$S_{pk}$} & \multicolumn{1}{c}{$S$} &\multicolumn{1}{c}{$V_{pk}$} & \multicolumn{1}{c}{$V_{\text{Range}}$} \\
			 &        & \multicolumn{1}{c}{\bf $h$~~~$m$~~~$s$}& \multicolumn{1}{c}{\bf $^\circ$~~~$\prime$~~~$\prime\prime$} & \multicolumn{1}{c}{(mJy)} & \multicolumn{1}{c}{(mJy~\kms)}    & \multicolumn{1}{c}{(\kms)}  & \multicolumn{1}{c}{(\kms)} \\ \midrule
			1 & M1  &   13 05 28.0 & $-$49 28 12.67 & 55$\pm2$ &  500$\pm40$ &   673   &     630--710    \\
			2 & M2  & 13 05 27.1   &   $-$49 28 07.39 & 5$\pm1$ & 250$\pm30$ & 490 & 420--560 \\ \hline
		\end{tabular} \label{tab:ngc4945_details}
	\end{center}
\end{table*}

\begin{table*} 
	\begin{center}
		\caption{NGC~253 36.2-GHz methanol emission details from the 2019 March H214 observations. All reported velocities are with respect to the barycentric reference frame. Labels here are based on those presented in \citet{Ellingsen+17b}. Uncertainties in the reported flux density values result from the random errors in fitting using the {\sc miriad} task imfit. }
		\begin{tabular}{@{}cllllccc@{}}
			\toprule
			 \multicolumn{1}{c}{ \bf$\#$} & \multicolumn{1}{c}{Label}  &\multicolumn{1}{c}{RA (J2000)}     & \multicolumn{1}{c}{Dec (J2000)} & \multicolumn{1}{c}{$S_{pk}$} & \multicolumn{1}{c}{$S$} &\multicolumn{1}{c}{$V_{pk}$} & \multicolumn{1}{c}{$V_{\text{Range}}$} \\
			 &        & \multicolumn{1}{c}{\bf $h$~~~$m$~~~$s$}& \multicolumn{1}{c}{\bf $^\circ$~~~$\prime$~~~$\prime\prime$} & \multicolumn{1}{c}{(mJy)} & \multicolumn{1}{c}{(mJy~\kms)}    & \multicolumn{1}{c}{(\kms)}  & \multicolumn{1}{c}{(\kms)} \\ \midrule
			1 & MM6 & 00 47 33.9  &  $-$25 17 11.65   &  25$\pm2$  &  605.3$\pm50$ & 207 & 160--260 \\
			 2 & MM5/MM4 & 00 47 33.7  &  $-$25 17 12.39   &  24$\pm1$  &  655.9$\pm50$ & 195 & 140--250 \\
			 3 & - & 00 47 32.8  &  $-$25 17 22.19   &  5$\pm2$ & 75.5$\pm20$  & 300 & 255--345 \\ 
			 4 & - & 00 47 32.3  & $-$25 17 19.06 &  5$\pm2$ & 69.6$\pm20$ & 336 & 300--360 \\ 
			 5 & MM1/MM2 &  00 47 32.0   & $-$25 17 28.98 & 30.0$\pm4$ & 1100$\pm60$ & 318 & 260--400 \\ \hline
		\end{tabular} \label{tab:ngc253_details}
	\end{center}
\end{table*}

\subsection{Data Reduction} \label{sec:data_reduction}

Data were reduced with {\sc miriad} using standard techniques for the reduction of 7-mm ATCA spectral line data. The data were corrected for atmospheric opacity and we estimate the absolute flux density calibration to be accurate to better than 30\%. For targets with continuum source detections, multiple iterations of phase-only self-calibration was performed. Any potential spectral line emission was isolated from the data by using the uvlin {\sc miriad} task. This task subtracts any potential continuum emission by estimating the correlated flux density on each baseline via polynomial interpolation based on line-free spectral channels. The velocity ranges and channels widths for our spectral line cubes varied depending on the source and CABB configuration/frequency setup that was utilised for the observations. When searching for spectral line emission, the data were resampled using many different channel widths, in order to appropriately identify both broad and narrow emission. Typical channel widths searched were 1, 3, 8, 10 and 20~\kms\ as the native spectral-resolution allows (dependent on the CABB configuration). Spectral line cubes are produced with a cell (pixel) sizes of $1\times1$ arcseconds and a Brigg's visibility weighting robustness parameter of 1. Details of the final spectral line cubes are listed in Table \ref{tab:cubes}, with image cube noise levels reported for a consistent channel width of 10~\kms\ for ease of comparison.

\subsubsection{Identification of maser sources}

Spectral line cubes were visually inspected for strong maser emission. The DUCHAMP source finder software \citep{Whiting12} was then used to search each cube with a $3\sigma$ threshold. Such a low threshold produces many false positives ($\sim$100), however the vast majority of these are on the outer edges of our images (which are typically $256''\times256''$). Any candidate within 2 times the FWHM of the primary beam (corresponding to a radius of $\sim$84$''$ from the pointing centre at 36.2-GHz) were manually inspected to determine if they were potential maser components. Generally with data of reasonable quality, very few false positives (less than two and typically none) were present within this central region of the image.

\begin{figure*}
	\begin{minipage}[h]{0.99\linewidth}
		\centering
		\includegraphics[scale=0.43]{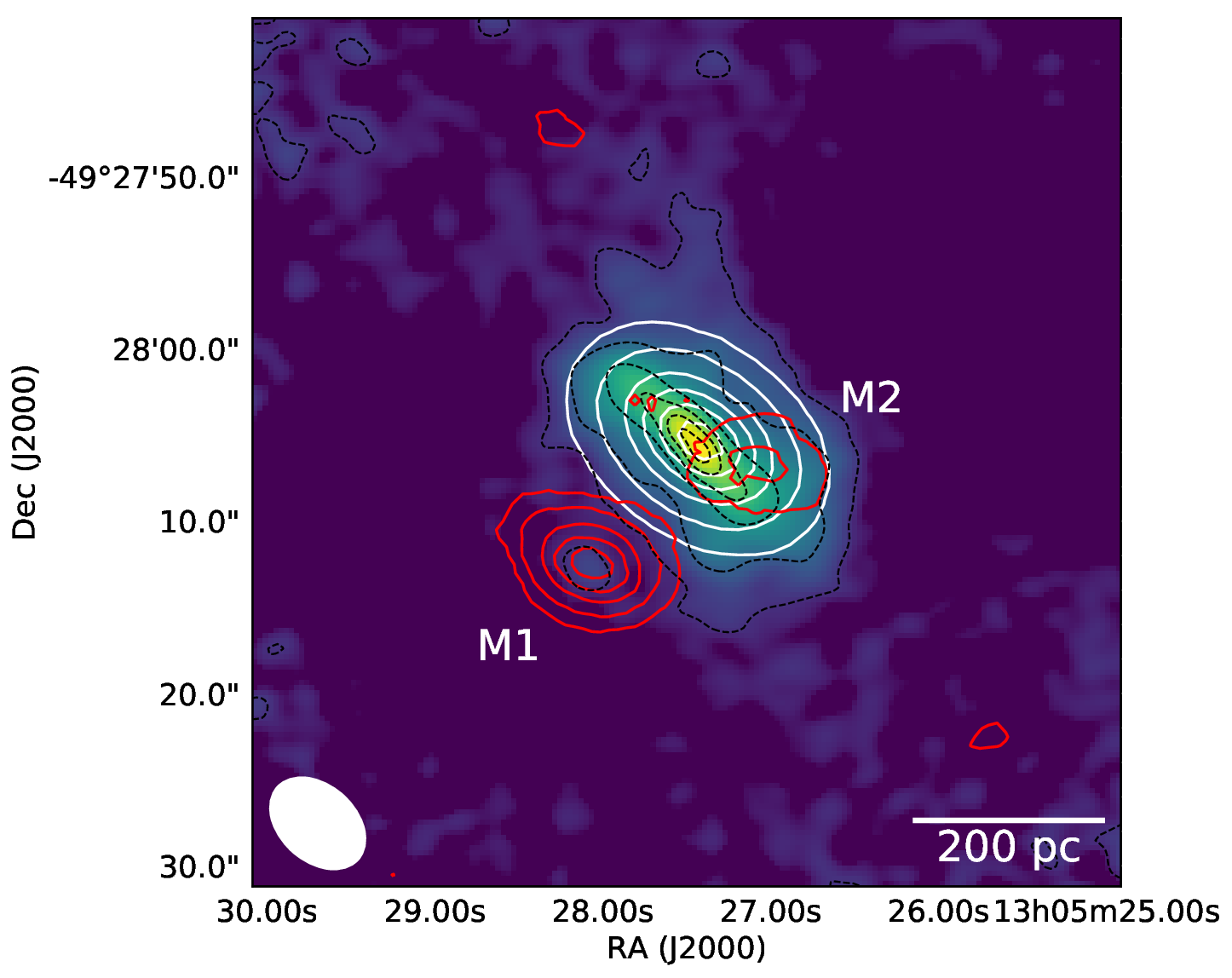}
	\end{minipage}
	\begin{minipage}[h]{\linewidth}
		\centering
		\includegraphics[scale=0.65]{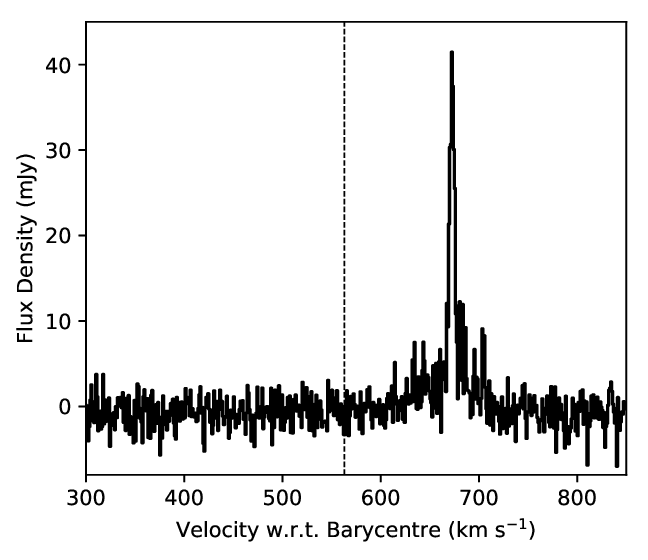}
		\includegraphics[scale=0.65]{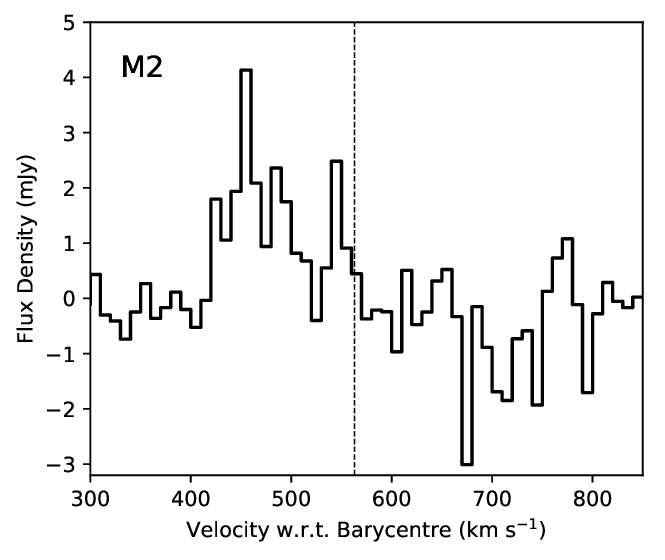}
	\end{minipage}
	\caption{Top: NGC~4945 integrated 36.2-GHz methanol emission (red contours at 10, 25, 50, 70 and 90 per cent of peak 560~mJy~\kms~beam$^{-1}$) and 7-mm continuum emission (white contours at 2, 10, 30, 50, 70, 90 per cent of peak 290~mJy~beam$^{-1}$) with  background colour map and black dashed contours of CS $J=2 \rightarrow 1$ integrated intensity from \citet{Henkel+18} (2\%, 10\%, 30\%, 50\%, 70\%, and 90\% of the peak of 11.5 Jy\,km\,s$^{-1}$ beam$^{-1}$). Methanol and continuum emission data extracted from the 2019 March spectral line cubes with a synthesised beam size of $6\farcs3\times4\farcs4$ (white ellipse). Bottom left: Spectrum of 36.2-GHz methanol taken from the location of peak emission at M1 (channel width of 1~\kms). The vertical dashed line indicates the systemic velocity of NGC\,4945 \citep{Chou+07}. Bottom right: Spectrum of 36.2-GHz methanol emission taken from the location of peak emission in region M2 (channel width of 12~\kms).}
	\label{fig:ngc4945_mar2019}
\end{figure*}

\section{Results} \label{sec:results}

This paper contains observations from three different projects, two involving searching for new extragalactic class I methanol maser sources (C2879 and C3263), and one studying known maser hosts (C3167). In order to facilitate easier reading we will separate the results based on project goal.

\begin{table*} 
	\begin{center}
		\caption{Details for the continuum emission from all sources. The horizontal line separates the maser search targets from the new (2019 March) H214 observations of NGC~253 and NGC~4945 (above and below respectively). The flux density values for NGC~1097 (denoted by an asterisk) have a higher level of uncertainty due to less accurate flux density calibration (see Section \ref{sec:2018_nov_dec}). For sources with no detected continuum source, the location of the pointing centre and a 5$\sigma$ upper limits on peak flux density are provided. Uncertainties in the reported flux density values result from the random errors in fitting using the {\sc miriad} task imfit.}
		\begin{tabular}{@{}lclll@{}}
			\toprule
			\multicolumn{1}{c}{Target Source}  & \multicolumn{1}{c}{Right Ascension}  &\multicolumn{1}{c}{Declination}     & \multicolumn{1}{c}{$S_{pk}$} & \multicolumn{1}{c}{$S$}  \\
			\multicolumn{1}{c}{} &   \multicolumn{1}{c}{\bf $h$~~~$m$~~~$s$}& \multicolumn{1}{c}{\bf $^\circ$~~~$\prime$~~~$\prime\prime$} &\multicolumn{1}{c}{(mJy~beam$^{-1}$)}  &\multicolumn{1}{c}{(mJy)}    \\ \midrule
			ESO\,269-g012             & 12 56 40.5 & --46 55 34.00 & $<2.3$ & - \\
			IC\,2560                 & 10 16 18.7 & --33 33 50.00 & $<1.8$ & -  \\
			IRAS\,11506-3851         & 11 53 11.8 & --39 07 49.18 & 10.0$\pm0.2$ & 10.7$\pm0.3$ \\
			NGC\,5793                & 14 59 24.8 & --16 41 36.70 & 18.8$\pm0.1$ & 19.3$\pm0.2$ \\
			ESO\,558-g009           & 07 04 21.0 &  --21 35 19.20 & $<0.6$ & - \\
			IRAS\,01417+1651         & 01 44 30.5 & +17 06 05.00 & $<0.7$ & - \\
			Circinus               & 14 13 10.0 & --65 20 20.74 & 56.6$\pm0.6$ & 63$\pm1$ \\
			IIzw96                     & 20 57 24.4 & +17 07 40.47 & 5.8$\pm0.2$ & 5.8$\pm0.3$ \\
			IRAS\,10173+0829       & 10 20 00.2 & +08 13 32.07 & 7.1$\pm0.9$ & 8$\pm2$ \\
			NGC\,1097             & 02 46 18.9 & -30 16 28.69 & 8.1$\pm0.1$* & 8.4$\pm0.1$* \\
			NGC\,1792             & 05 05 14.5 & --37 58 49.87 & 4.4$\pm0.1$ & 4.4$\pm0.1$ \\
			NGC\,1808             & 05 07 42.3 & --37 30 45.64 & 4.5$\pm0.1$ & 6.2$\pm0.4$ \\  \hline 
			NGC\,253             & 00 47 33.2  &  $-$25 17 17.77   & 214.3$\pm8$   & 335.7$\pm20$ \\ 
			NGC\,4945             & 13 05 27.4  & $-$49 28 05.43 & 296$\pm3$ & 430.8$\pm7$ \\  \hline                  
		\end{tabular} \label{tab:continuum}
	\end{center}
\end{table*} 

\subsection{Project C3167: the methanol masers of NGC~253 and NGC~4945}

\subsubsection{NGC~4945} 
The previously reported bright 36.2-GHz methanol maser offset to the south-east from the nucleus of NGC~4945 was readily detected in the latest 2019 March epoch observation (M1 in Figure \ref{fig:ngc4945_mar2019}). 
In addition to this primary region, we identify another region of 36.2-GHz emission south-west of the galactic centre. This emission is offset by approximately 3 arcseconds to the south-west of the galactic nucleus (M2 in Figure \ref{fig:ngc4945_mar2019}). This location corresponds to the south-west component of 36.4-GHz HC$_3$N and CS ($1-0$) emission reported in \citet{McCarthy+18c} and the methanol emission covers the same velocity range. When compared to the primary region, this new emission has a much lower peak flux density and a significantly broader spectral profile. This emission is clearly visible in the data prior to any self-calibration, indicating it is not an artefact of this process.

7-mm continuum emission was detected toward the nucleus of NGC~4945. As in \citet{McCarthy+18c}, this continuum measurement was obtained using the line-free data in the 36.2-GHz zoom band. The measured flux density is in agreement with the previously reported measurement. The integrated flux density agrees with the modelled 7-mm continuum values from \citet{Bendo+16}, which predicts that 75 percent of this 7-mm continuum emission is from free-free emission. 

The flux density values for the methanol and 7-mm continuum emission toward NGC~4945 are recorded in Table~\ref{tab:ngc4945_details} and \ref{tab:continuum} respectively.. 

\subsubsection{NGC~253} 

We detect the same 36.2-GHz methanol maser regions toward NGC~253 as have been reported by \citet{Ellingsen+14,Ellingsen+17b}, though the resolution of the new observations does not allow us to resolve all of the individual features at the locations of the brightest emission. 
We also identify two tentative new components of methanol emission (see Figure \ref{fig:ngc253_mar2019}) not formerly detected by \citet{Ellingsen+17b} or \citet{Gorski+17}. These components are only approximately 4$\sigma$ detections, however, they are located at the same positions and velocities as emission in the 87.9 GHz transition of HNCO emission \citep[location C and D in figure 1 of][]{McCarthy+18b}. The previous 36.2-GHz ATCA observations of NGC~253 by \citet{Ellingsen+17b} have a high enough sensitivity (1.1 mJy~beam$^{-1}$ in a 3~\kms\ channel), however, their observations make use of more extended array configurations. Both components have sufficiently high peak flux densities ($\sim$5~mJy) that they should have been detected by previous observations of NGC~253.

7-mm continuum emission was detected toward the nucleus of NGC~253. Continuum emission was extracted from the line-free data in the 36.2-GHz zoom band, flux density values for this continuum emission, along with those for the maser components are recorded in Table \ref{tab:ngc253_details}.

\subsection{Projects C2879 \& C3263: searching for masers} \label{sec:results_c2879_c3263}

No additional extragalactic methanol maser sources have been detected in any of the targets of these projects and RMS noise levels for the 36.2-GHz class~I and 37.7-GHz class II spectral line cubes are listed in Table \ref{tab:cubes}. Results for NGC~1097 have a higher uncertainty due to flux-calibration issues, as outlined in Section \ref{sec:2018_nov_dec}.

7-mm continuum emission was detected toward the centre of many of these galaxies (see Table \ref{tab:continuum}), many of which have no previously reported values in the literature.

\section{Discussion}

\subsection{New methanol emission toward NGC~4945} \label{sec:new_meth_ngc4945}

The new methanol emission from region M2 shares overlapping position and velocities with the south-western CS (1--0) and HC$_3$N (4--3) emission reported in \citet{McCarthy+18c}. This increases our confidence that despite a relatively low-SNR, the emission at this location is indeed real. Before we can discuss what this new region of methanol emission means for our understanding of extragalactic class~I methanol masers, we must first justify that it is reasonable to consider this emission the result of maser processes.

While not as narrow as the linewidths seen toward the primary maser region (M1), M2 appears to consist of two or more components with FWHM of $\sim$20~\kms. These linewidths are much narrower than those observed from thermal emission toward the same regions \citep[FWHM of $>60$~\kms;][]{Henkel+18, McCarthy+18c}, and are comparable to the masing regions toward NGC~253 \citep{Ellingsen+17b}. Similarly to the main 36.2-GHz methanol masing region, no methanol from the ground state 48 GHz transition has been reported toward M2 \citep{McCarthy+18c}. 

When investigating the emission at M2 in the three epochs of observation originally presented in \citet{McCarthy+18c}, it should be noted that all were made in relatively compact ATCA array configurations (2016 Aug, 2017 Jun and Oct; see Table \ref{tab:variability}). Looking at these three array configurations we see the integrated flux density of M2 increasing as the array configuration gets more compact. This relationship may indicate thermal emission from this region, however, there are a few important caveats to this interpretation. The first is that the detections of emission from M2 in these epochs is very marginal, with only a channel or two above the noise level (see Figure \ref{fig:ngc4945_central_spectra}). This means the extracted integrated flux density values from these locations with marginal detections is less reliable. Secondly, a similar relationship is observed toward M1, where integrated flux density increases as they array becomes more compact. Finally, it is hard to determine the true integrated flux density for M2 from the 2017 July H75 observations, as the emission from the primary component M1 is also captured within the larger synthesised beam.

The emission from this new methanol region also displays evidence of variability. This is the best evidence that the emission does not result from a thermal process. This variability is discussed in more detail in Section \ref{sec:variability}, as it is the first compelling evidence for variability in extragalactic class~I methanol masers.

Due to the broad spectral profile of the emission and the larger velocity range, the integrated flux density of M2 is only approximately 50\% lower than the primary masing region, despite a much lower peak flux density (by a factor of $\sim$10). This integrated intensity is approximately 5 orders of magnitude higher than that of a typical Galactic style 36.2-GHz emission, and over an order of magnitude (a factor of $\sim$40) higher than the combined 36.2-GHz emission from the Milky Way CMZ \citep{Yusef-Zadeh+13}. As this region is unresolved by our synthesised beam ($6.3\times4.4$ arcseconds), we can put an upper limit on its angular size. This angular size corresponds to a linear size of $113 \times 79$ pc at the 3.7~Mpc distance of NGC~4945 (likely a vast over-estimate based on variability timescale, see Section \ref{sec:variability}). This implies we have over 40 times the 36.2-GHz emission coming from an area on the sky which is at most 30\% larger than the CMZ would be at this distance. 

Considering all these factors together, we conclude that the most reasonable explanation is that this emission is masing. \citet{McCarthy+18c}  also put an upper limit of 6 mJy on the presence of any 44-GHz methanol emission from NGC~4945. This means the 36.2-GHz methanol emission from this region is unlikely to be resulting from the cumulative effect of Galactic-style class~I masers, as otherwise we would expect to also see evidence of 44.1-GHz maser emission \citep{Voronkov+14}.

%

\subsection{Tentative new methanol emission in NGC~253}

\begin{figure*}
	\begin{minipage}[h]{0.5\linewidth}
		\centering
		\includegraphics[scale=0.70]{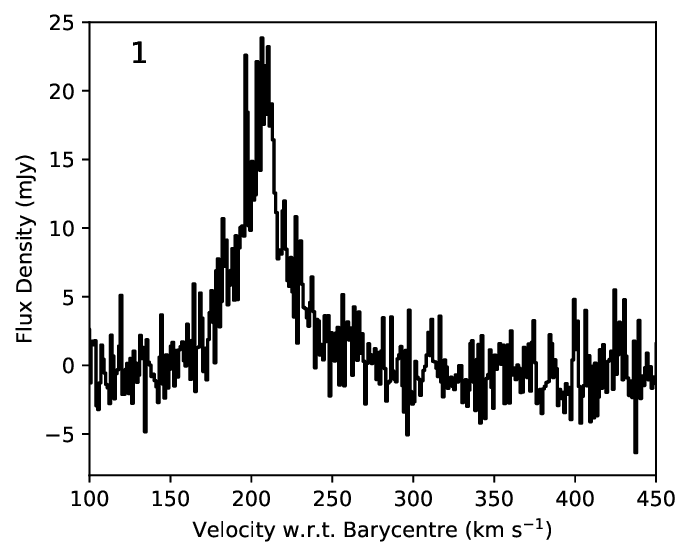}
	\end{minipage}
	\begin{minipage}[h]{0.49\linewidth}
		\centering
		\includegraphics[scale=0.70]{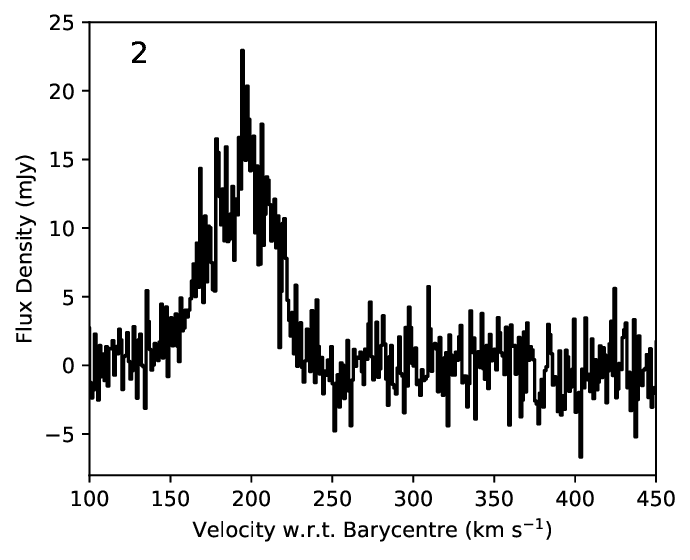}
	\end{minipage}
	\begin{minipage}[h]{0.5\linewidth}
	\centering
	\includegraphics[scale=0.37]{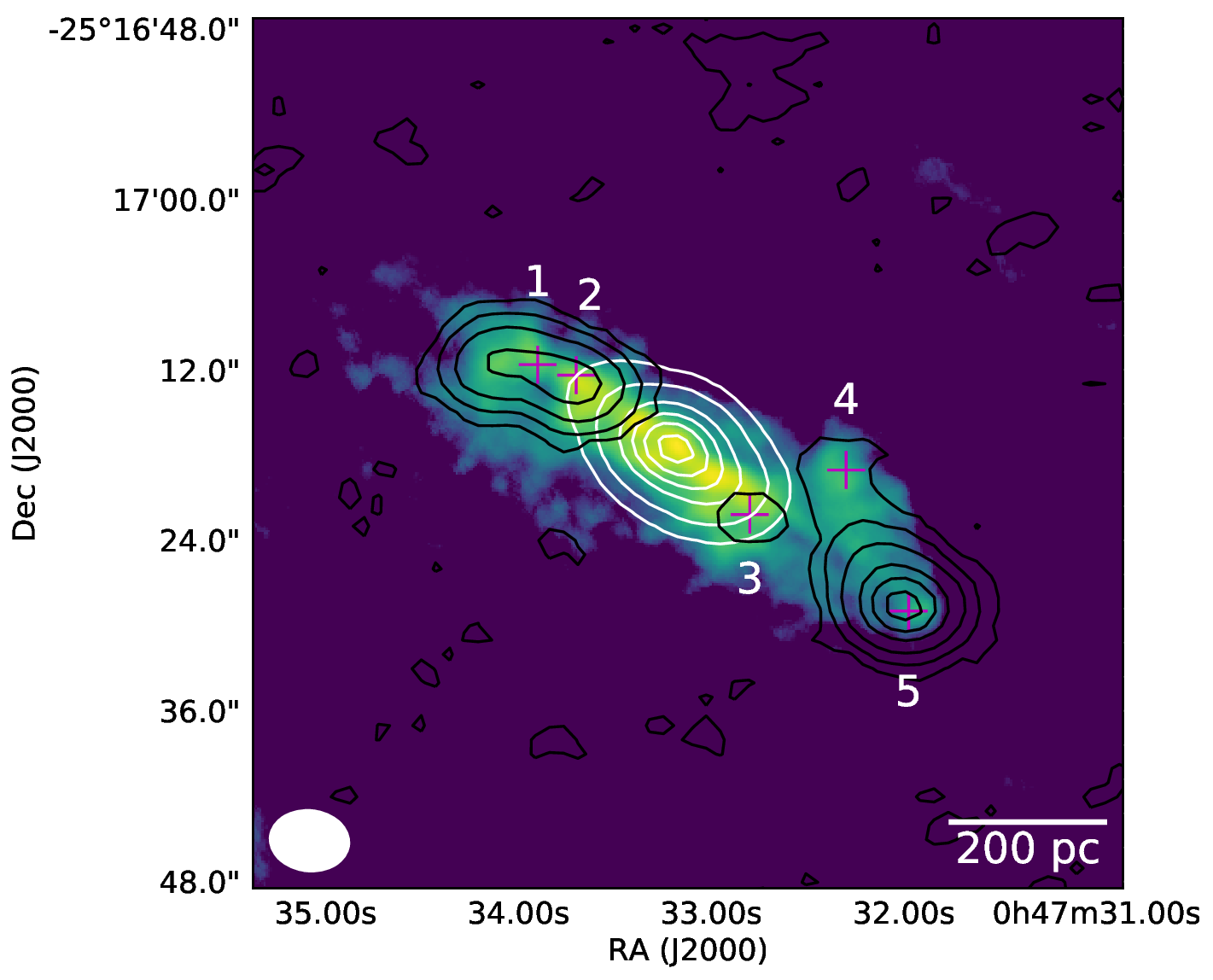}
    \end{minipage}
    \begin{minipage}[h]{0.49\linewidth}
	\centering
	\includegraphics[scale=0.70]{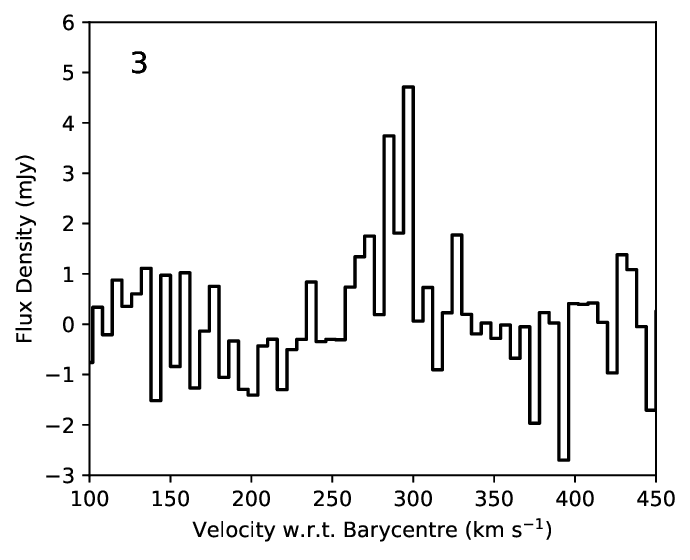}
    \end{minipage}
	\begin{minipage}[h]{0.5\linewidth}
	\centering
	\includegraphics[scale=0.70]{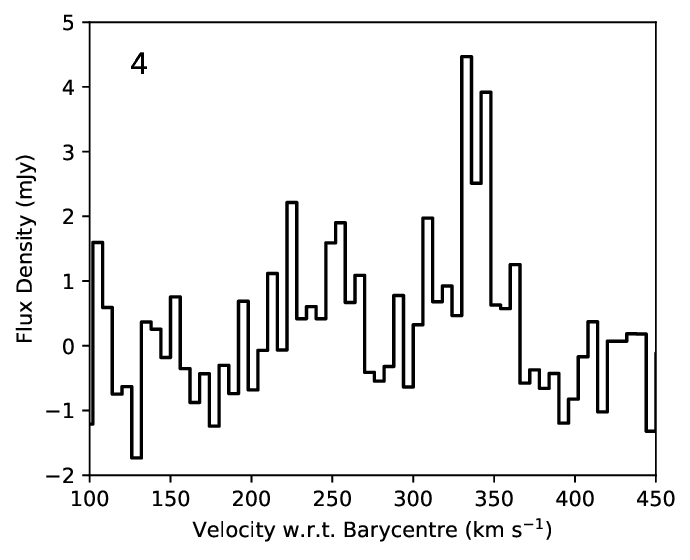}
    \end{minipage}
    \begin{minipage}[h]{0.49\linewidth}
	\centering
	\includegraphics[scale=0.70]{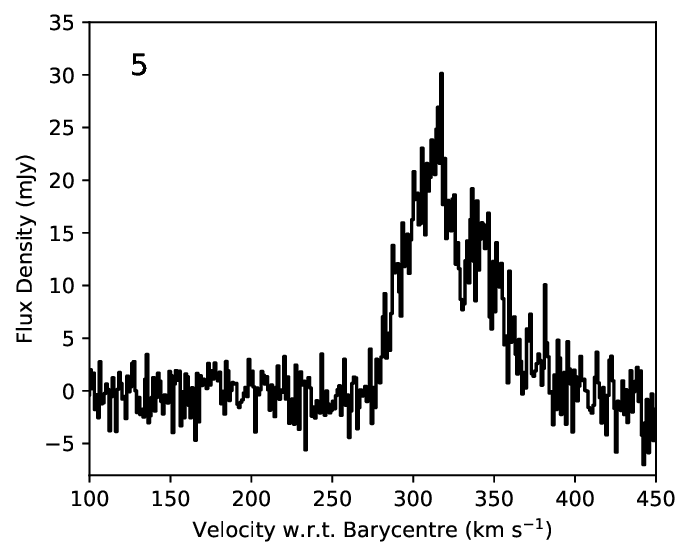}
    \end{minipage}
	\caption{All data in this figure is from the 2019 March epoch, in the ATCA H214 array configuration. Field image: NGC~253 integrated 36.2-GHz methanol emission (black contours 2.5, 10, 25, 50, 70, and 90 per cent of the 1364 mJy\,\kms ~beam$^{-1}$ peak) and 7-mm continuum emission (white contours 2.5, 10, 30, 50, 70, and 90 per cent of the 215 mJy~beam$^{-1}$ peak) with background image of integrated CO $J = 2 \rightarrow 1$ emission from \citet{Sakamoto+11} on a logarithmic scale. Magenta plus signs indicate the peak components of the methanol emission and have been labelled 1 through 5 moving from high right ascension to low. Spectra: Surrounding spectra from the 36.2-GHz spectral line cube at the location of each magenta plus sign. Synthesised beam size for our observations is $5\farcs6 \times 4\farcs3$ (white ellipse). Spectra 1, 2 and 5 have a channel width of 1~\kms, while 3 and 4 have 10~\kms\ channel width in order to better show the velocity range of the weaker emission from these regions.}
	\label{fig:ngc253_mar2019}
\end{figure*}

The angular resolution of our observations does not allow us to separate the methanol emission into the individual components as reported in previous interferometeric studies \citep{Ellingsen+17b, Gorski+17, Chen+18}. However, despite this we observe two new components of emission (labelled 3 and 4 in Figure \ref{fig:ngc253_mar2019}). These components are only $4\sigma$ detections, however, their location and velocity strongly correlate with the components of HNCO emission reported in \citet{McCarthy+18b}. \citet{Ellingsen+17b} observe a close correlation between the location of 36.2-GHz methanol masers and thermal emission from HNCO, a low-velocity shock tracer. This indicates that despite being a marginal detection, these components are likely real.

Comparison of our new observations with past ATCA observations of the 36.2-GHz methanol masers toward NGC~253 \citep{Ellingsen+14, Ellingsen+17b}, the whole spectrum (combining all regions) is more similar to what is reported in the original detection paper. The `intermediate resolution' observations displayed in figure 1 of \citet{Ellingsen+17b} shows significantly higher peak flux density values then what we see from our observations or their past observations \citep{Ellingsen+14}. However, upon re-analysis of the data, this appears to be an error (with flux density values out by a factor of 2) in how the spectrum has been extracted from their `intermediate resolution' spectral line cube (and it does not affect the spectra taken from the combined array cube in figure 2 of \citet{Ellingsen+17b}), and the data is consistent across all 3 epochs.


\begin{figure}
	\includegraphics[width=\linewidth]{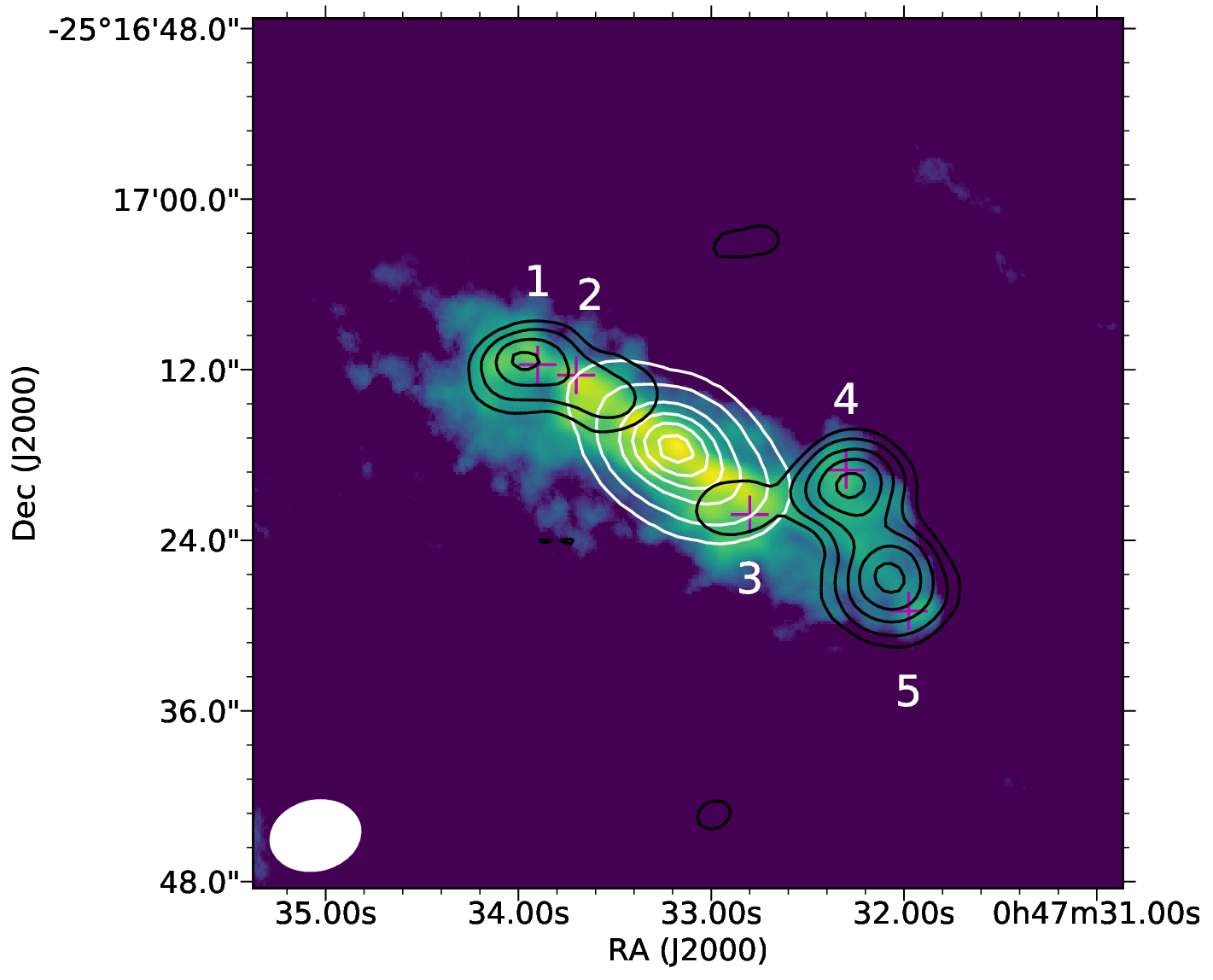}
	\caption{NGC~253 integrated 87.9~GHz HNCO emission (black contours 15, 30, 50, 70, and 90 per cent of the 1.9 Jy\,\kms ~beam$^{-1}$) and the 7-mm continuum emission (white contours 2.5, 10, 30, 50, 70, and 90 per cent of the 215~mJy~beam$^{-1}$ peak) with background image of integrated CO $J = 2 \rightarrow 1$ emission from \citet{Sakamoto+11} on a logarithmic scale. Magenta plus signs indicate the peak components of the 36.2-GHz methanol emission and have been labelled 1 through 5 moving from high right ascension to low. Synthesised beam for the 87.9-GHz HNCO emission is approximately $6\farcs4 \times 4\farcs9$.}
	\label{fig:meth36_hnco84}
\end{figure}

\begin{figure}
	\includegraphics[width=\linewidth]{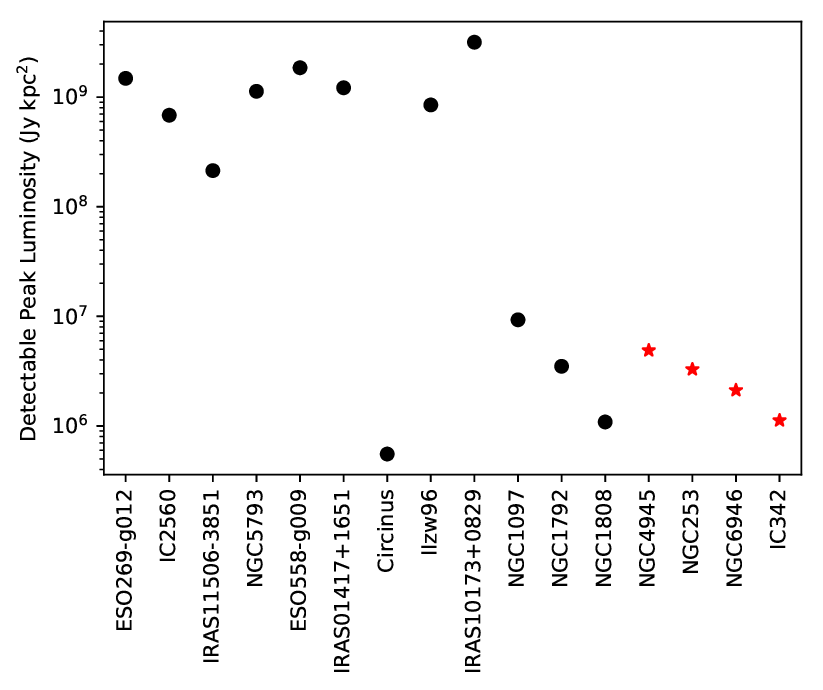}
	\caption{Black dots indicate the detectable peak emission levels for each target source in the maser search projects. Red stars represent the peak luminosity levels in each of the four sources with detected 36.2-GHz class~I methanol masers. All peak emission levels are given for a 10~\kms\ velocity channel. Peak or integrated luminosity values for this figure (and throughout the rest of the paper) are given with units of Jy~kpc$^2$ (or Jy~\kms~kpc$^2$ for integrated luminosity), which is the flux-density value multiplied by the squared distance to the source (in kpc).}
	\label{fig:sensitivity}
\end{figure}

\begin{figure*}
	\begin{minipage}[h]{0.5\linewidth}
		\centering
		\includegraphics[scale=0.70]{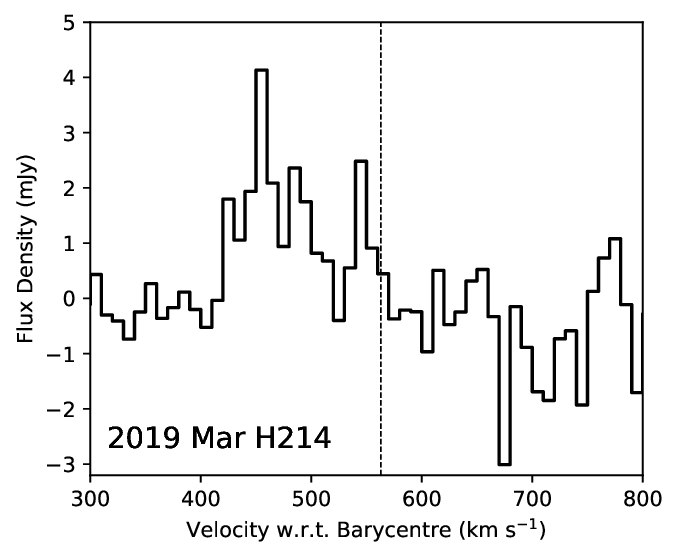}
	\end{minipage}
	\begin{minipage}[h]{0.49\linewidth}
	\centering
	\includegraphics[scale=0.70]{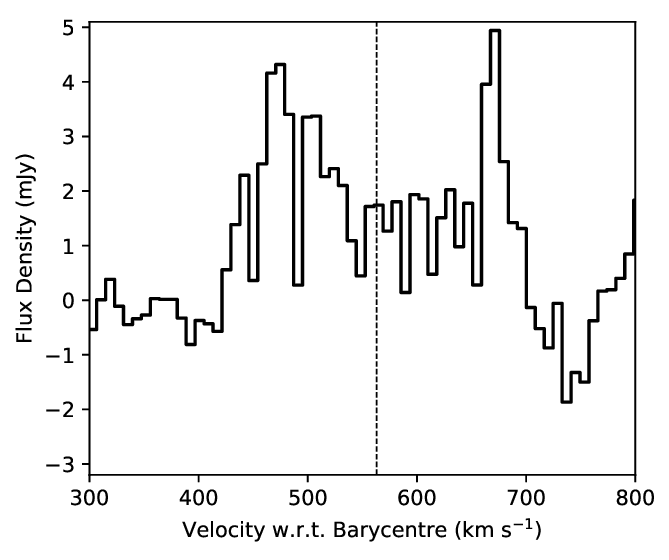}
    \end{minipage}
	\begin{minipage}[h]{0.50\linewidth}
		\centering
		\includegraphics[scale=0.70]{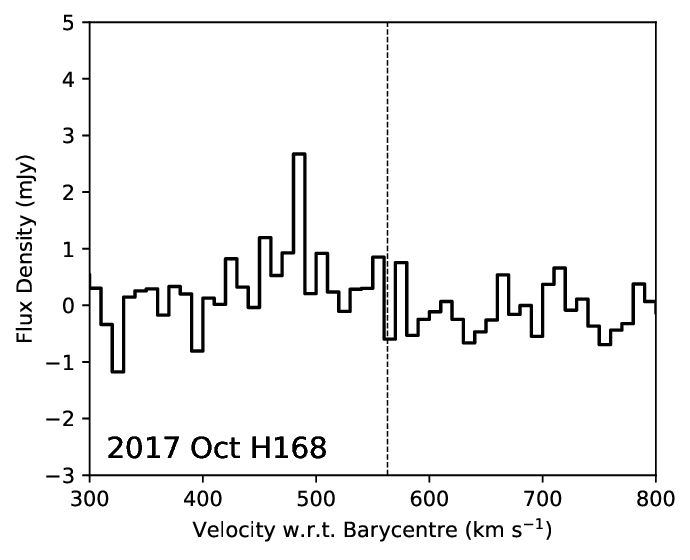}
	\end{minipage}
	\begin{minipage}[h]{0.49\linewidth}
	\centering
	\includegraphics[scale=0.70]{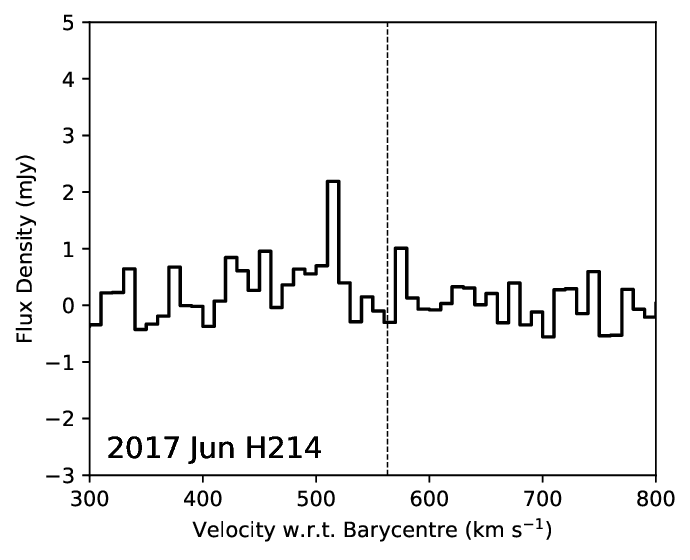}
    \end{minipage}
	\begin{minipage}[h]{\linewidth}
	\centering
	\includegraphics[scale=0.70]{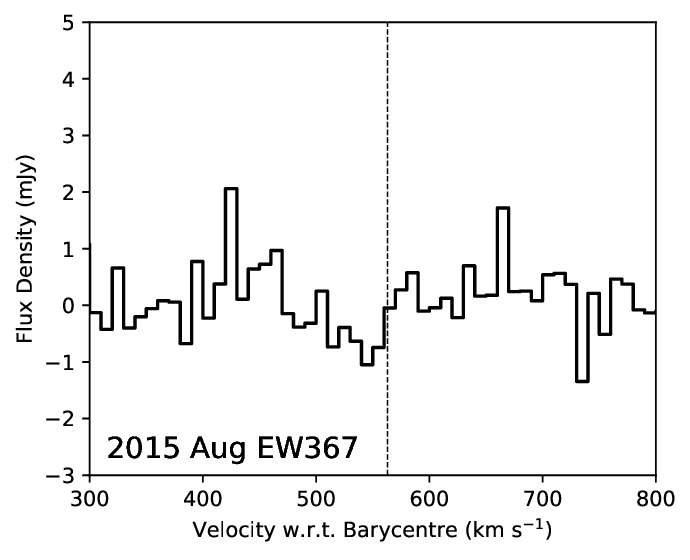}
    \end{minipage}
	\caption{Spectra of 36.2-GHz emission at M2 toward NGC~4945 from each individual observation epoch (two epochs presented in this paper and the three from \citet{McCarthy+18c}). Vertical scales have all been set to the same range for ease of comparison. Note for the 2017 Jul H75 spectrum, the emission at $\sim$670\kms\ is due to the larger synthesised beam capturing emission from the main maser component in NGC~4945. The vertical dashed line indicates the systemic velocity of NGC\,4945 \citep{Chou+07}.}
	\label{fig:ngc4945_central_spectra}
\end{figure*}

\begin{figure}
	\includegraphics[width=\linewidth]{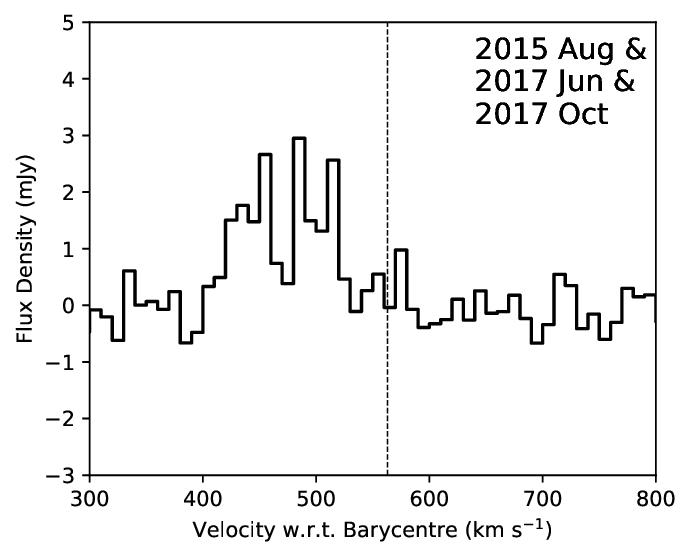}
	\caption{The spectrum from the M2 methanol maser region in NGC~4945 using the combined data from the three epochs reported in \citet{McCarthy+18c}.}
	\label{fig:3epoch}
\end{figure}

\subsection{Evidence for variability in extragalactic methanol masers} \label{sec:variability}

We currently have five epochs of observations of the 36.2-GHz methanol transition toward NGC~4945, three of which were originally reported in \citet{McCarthy+17} and \cite{McCarthy+18c}, with the remaining 2 reported in this paper. This allows us to compare the methanol emission over a time baseline of approximately 3.5 years. Tabulated flux density values for the primary methanol maser component (M1), newly discovered central region (M2) and 7-mm continuum emission can be found in Table \ref{tab:variability}.

The detection of the new emission at M2 in NGC~4945 is the first strong evidence for variability in an extragalactic class I methanol maser source. This is most evident when comparing the two observations made using the H214 array configuration, one in 2019 March and the other in 2017 June (see Figure \ref{fig:ngc4945_central_spectra} for spectra and Table \ref{tab:variability} for flux density values). Due to the difficulties of comparing emission observed with different array configurations, the comparison of these two H214 epochs will be the main focus of this section. We see a 330 percent increase in integrated flux density across the velocity range of the new component (420--560~\kms) when comparing these two epochs (see Table \ref{tab:variability}). We also see evidence of this emission in the much lower resolution (synthesised beam size of $26.6 \times 12.0$ arcseconds) H75 array data from 2017 July, which is bracketed by the higher resolution observations in both 2017 June and October, neither of which show significant emission. This may indicate that emission from this region was much more diffuse during this period, with it becoming significantly more compact sometime prior to the most recent observation. However, if we assume a time-scale of 21 months for the variability (based on the time period between the H214 observations), this puts an upper limit on the size of the emitting region of 21 light-months ($\sim$0.53\,pc). At a distance of 3.7~Mpc, this corresponds to an angular size of 0.03 arcseconds, meaning the emission would be compact enough to be detectable on any ATCA baseline.  It is therefore more likely that the change in luminosity of the region is instead related to an excitation change. Combining all data from the three intermediate resolution observations reported in \citet{McCarthy+18c} (see Figure \ref{fig:3epoch}), we can clearly see this emission at M2 (despite the the individual epochs showing no clear evidence) with a peak flux density of $\sim3$~mJy and integrated flux density of 185 mJy~\kms. This further supports the interpretation that the variability is a result of an excitation change, rather than the emission simply being more diffuse during these earlier epochs. 

The angular size of 0.03 arcseconds also allows us to put a lower limit of $\sim$6500\,K on the brightness temperature of the emission from M2 based on the peak spectral channel (channel width 10~\kms), or $>3\times10^5$\,K integrated over a 140~\kms\ velocity range. This brightness temperature provides further evidence that this emission is the result of a non-thermal process, as discussed in Section~\ref{sec:new_meth_ngc4945}.


The brighter region to the south-east of the nucleus (M1 in Figure \ref{fig:ngc4945_mar2019}) also shows some evidence of variability. The multiple spectral components reported in \citet{McCarthy+18c} are all present without any measurable shift in their velocities. When directly comparing the emission from the south-eastern bright region to the 2017 June observations in the same array configuration (H214), we observe a decrease in peak flux density of $\sim$6 percent and increase in integrated flux density of $\sim$30 percent. The integrated flux density of this primary region is also significantly higher ($\sim$15 percent) than that observed using a more compact array configuration (H168) in 2017 October.

Our previous 2017 June H214 observations, when compared to other 7-mm observations toward NGC~4945, had lower continuum values than expected \citep{McCarthy+18c}. However, here we observe the same values of peak and integrated flux density for the 7-mm continuum source as reported from the previous H214 observations of NGC~4945 (see Table \ref{tab:variability} for values). This suggests that instead of the 2017 June continuum levels being anomalously low, the 2015 August EW367 continuum levels may be higher than expected. This is likely related to the poor uv-coverage (highly elongated beam) of these observations preventing accurate imaging (flux density values are extracted from images using the imfit miriad task). The consistency of the 7-mm continuum measurements between these two H214 observations increases the confidence in our flux density calibration, which strengthens our comparisons of the maser emission between these two epochs. 

It can be difficult to measure variability in the weak millimetre spectral line emission from these extragalactic sources. It is difficult to isolate what is variability in the source, and what is variability caused by differing array configurations, weather and uv-coverage between observations. With that said, the increased integrated flux density observed from M2 in our recent observations (2019 March) when compared to past observations (in particular the 2017 Jun H214 epoch) is far too large to be the result of weather or systematic effects.

The tentative new components detected toward NGC~253 also may indicate variability in this emission. The previous observations of NGC~253 at 36.2-GHz \citep[][]{Ellingsen+17b} have appropriate sensitivity to detect these components, however, these observations use a higher-resolution east-west oriented array without complete uv-coverage. The resulting synthesised beam of those observations makes determining component separation along the north-east to south-west axis difficult, which is required to resolve the individual south-western components.

The mechanism driving such significant variation in these masers is mostly a mystery, especially because at this stage we do not understand the nature of the variability. It could be that it is a periodic phenomenon, or instead we may be observing the creation of new masers at these locations which will eventually stabilise. In Galactic sources, class~I methanol masers are observed to exist for long periods of time, remaining relatively stable \citep{Kurtz+04}. This stability is consistent with the expected variability time-scales for saturated class~I masers, where the low-velocity shocks have long relaxation times \citep[15 years;][]{Leurini+16}. However, it should be noted that long-term monitoring of class~I methanol maser sources has not been undertaken to date, and our understanding of variability in these Galactic class~I masers is lacking. Another collisionally pumped maser species is water, which is observed to be highly variable in both Galactic and extragalactic sources \citep[e.g.][]{Haschick+90, Felli+07,McCallum+07}. These water masers trace more energetic environments, with higher velocity shocks \citep{Hollenbach+13}. It may be that extragalactic class~I methanol masers also trace higher-energy environments than their Galactic counterparts, though this is largely speculation as all modelling of the class~I methanol lines is based on our understanding of Galactic sources \citep{Sobolev+07, McEwen+14, Leurini+16}. In order to confirm variability in these extragalactic class~I methanol masers and develop our understanding of the phenomenon, constant and consistent monitoring is required. This monitoring should ideally be made with the same spatial resolution in order to accurately quantify any variability observed toward these sources.

\begin{table*}[]
	\begin{tabular}{lllllc}
		\hline
		\multicolumn{1}{c}{Epoch} & \multicolumn{4}{c}{Flux density source (mJy)} & \multicolumn{1}{c}{Angular Scale} \\
		& \multicolumn{1}{c}{M1} & \multicolumn{1}{c}{M2} & \multicolumn{1}{c}{Continuum} & \multicolumn{1}{c}{Phase Cal.} & \multicolumn{1}{c}{Min/Max (arcsec).} \\ \hline
		2015 Aug (EW352) & 27.9 (289) & 2.1 (44) & 313 (414) & 3328 & 3.6/55.8 \\
		2017 Jun (H214) & 27.4 (305) & 2.3 (65) & 236 (348) & 1369 & 4.4/20.7 \\
		2017 Jul (H75) & 21.3 (450) & 4.4 (279) & 452 (447) & 935 & 12.0/55.8 \\
		2017 Oct (H168) & 30.4 (346) & 2.8 (71) & 385 (489) & 1141 & 5.0/27.9  \\
		2019 Mar (H214) & 25.7 (397) & 4.3 (215) & 246 (356) & 1110 & 4.4/20.7 \\ \hline
	\end{tabular} 
    \caption{Table of flux density values from each epoch for the emission at M1, M2 and the continuum for NGC~4945, including the phase calibrator for these observations. Peak values are given, with integrated values given in parentheses. The peak values at both region M1 and M2 are given with a channel size of 10~\kms. It should be noted that these flux density values have an uncertainty of approximately 30\% due to flux density calibration and continuum subtraction (see Section \ref{sec:data_reduction}). We have left the random fitting errors off these values in order to facilitate easier reading, in general these random errors are around~5\%.}
    \label{tab:variability}
\end{table*}

\subsection{Non-detection in Arp~220 follow-up}

The weather during the 2014 March observations, was quite poor (particularly on March 28). During the processing of the data from this epoch we found that certain imaging parameters were causing spurious emission to appear in the spectral line cubes of targets with relatively low on-source time (and therefore, high noise). This realisation prompted re-analysis of the Arp~220 data published in \citet{Chen+15}, as the data reported in that paper was from this observing session, and has not been confirmed in any subsequent observations. Upon reanalysing with more robust imaging parameters we find a similar spectral profile to that which is reported in \citet{Chen+15}, however, at much lower flux density than that reported ($\sim$6~mJy peak compared with $\sim$25 mJy). The RMS noise in the cube is 4.3 mJy, so we can not be confident that this is real emission. Additionally, we see no evidence for the reported 37.7-GHz emission \citep{Chen+15} in the re-analysed data. We have also reduced data from an ATCA follow-up observation in 2014 November (EW367 array configuration), with much longer integration times (4.8 hours). Analysis of these data does not show any 36.2-GHz methanol emission toward the central region of Arp 220 comparable to that reported in \citet{Chen+15}. Based on these follow up observations, we put a $5\sigma$ upper limit on emission from the 36.2-GHz methanol line of 3.5~mJy. A recent 36.2-GHz search by Henkel et al. (submitted) targetted Arp~220 with the Effelsberg 100m telescope and report no detection of 36.2-GHz emission (with an upper limit of 4.2~mJy).

\subsection{Sample discussion} \label{sec:sample}

When considering the non-detections toward most of our targets, it is important to note that the observations conducted under project C2879 were also responsible for discovering the first extragalactic methanol masers. Hence at the time the observations were made there was no information on the luminosity, nor likely environments to host extragalactic class I methanol masers. This results in many of the target sources not having a high enough sensitivity to detect methanol masers of comparable luminosity to those observed toward our known hosts (see Figure \ref{fig:sensitivity}). This was not a problem for the subsequent searches, which uses much higher sensitivity levels, based on the currently known extragalactic class~I maser sources.

Of the 12 sources for which we report 36.2-GHz methanol upper limits (C2879 and C3263 targets), only four have sensitivity levels adequate for detecting masers with the same luminosity as those detected toward NGC~4945, with the sensitivity of the majority over an order of magnitude too low (see Figure \ref{fig:sensitivity}). When combining the non-detections with detections for analysis of the sample, the majority of the time we will include only these sources with appropriate sensitivity levels. This means our primary sample for discussion will include: NGC~4945, NGC~253, NGC~6946, NGC~1068, NGC~1097, NGC~1792, NGC~1808, IC~342 and the Circinus Galaxy.

Considering the general properties of the galaxies with detections we see three common factors: They are all barred, spiral galaxies with elevated levels of star-formation toward their nucleus (when compared to the Milky Way). However, it is important to note that searches for extragalactic methanol masers have been biased toward these types of sources based on the first detections in NGC~253 and NGC~1068 \citep{Ellingsen+14,Wang+14}. As of yet no statistically complete search for these masers has been undertaken, and as such, it is hard to determine the most relevant properties for future target selection. The observations for project C3263, reported in this paper, used the above properties to target nearby southern sources. Despite that, the three sources observed so far as part of that project (NGC~1097, NGC~1792 and NGC~1808) have not shown any sign of methanol maser emission. The only other source observed with appropriate sensitivity is the Circinus Galaxy, which is a spiral galaxy with no bar, however it does show evidence of molecular outflows from its type 2 Seyfert AGN.

In order to determine the most efficient way, in terms of target selection and telescope time, to search for extragalactic 36.2-GHz methanol masers, we must determine the luminosity levels of a `typical' region. The currently reported extragalactic 36.2-GHz methanol masers (in NGC~4945, NGC~253, NGC~6946 and IC~342) all have varying peak and integrated luminosity levels. In this discussion we consider the peak and integrated flux density from a single emission component at our resolution (rather than the combined emission across the whole source) as these are what will be detected when searching using interferometers like the VLA or the ATCA. Peak luminosities (in a 10~\kms\ channel) for the brightest region in each source vary from $1.1\times10^6$~Jy~kpc$^2$ in IC~342 to $5.2 \times 10^6$~Jy~kpc$^2$ toward NGC~4945, with integrated luminosities ranging from $2.8 \times 10^7$~Jy\,\kms\,kpc$^2$ in IC~342 to $1.6 \times 10^8$~Jy\,\kms\,kpc$^2$ toward NGC~253. It must be noted that the flux density values for IC~342 and NGC~6946 are from observations with much higher spatial-resolution than the ATCA observations (VLA C-configuration) of NGC~4945 and NGC~253 ($\sim$1 arcsec compared to $\sim$5 arcsec). We calculate an average peak luminosity of $3.0 \times 10^6$~Jy~kpc$^2$ across these 4 sources. With a sensitivity of 0.5~mJy (in a 10~\kms\ channel), a 5$\sigma$ detection of a maser with this luminosity can be made out to a distance of $\sim$34~Mpc. Assuming the ability to self-calibrate off a continuum source, this sensitivity can be achieved with about 4 hours of on-source time with the ATCA in an intermediate resolution array configuration (H168 -- EW367) or 45 minutes of on-source time with VLA D-configuration (VLA D-configuration will allow detection of any extended emission, such as that seen toward NGC~253). A 10~\kms\ channel width is appropriate for searching for these masers as this will allow multiple, above noise level, consecutive channels across the FWHM of the spectral profile for the vast majority of known extragalactic 36.2~GHz methanol sources. The only exception to this is the spectral profile of the M1 region in NGC~4945, which has a FWHM of $\sim$20~\kms. Though this source has a much higher peak luminosity then the other sources, so the ability to image with narrower channel widths will still allow flexibility for detecting these sources. However, it should be noted that our analysis of optimal search parameters for these 36.2-GHz masers is based off a sample size of four sources, and as such is susceptible to small number statistics.



\subsection{Galaxy-wide star-formation as a predictor for class~I methanol masers}

Within our Galaxy, methanol masers (both class~I and II) provide useful sign-posts for the location of star-formation regions \citep{Caswell97, Kurtz+04,Voronkov+06,Breen+10a,Breen+11}. It was initially hypothesised that these extragalactic class~I masers may be useable in the same manner, identifying starburst galaxies at cosmologically interesting distances \citep{Chen+16}. However, that speculation was largely predicated on the more luminous emission reported toward Arp 220. Reanalysis showing no 36.2-GHz methanol emission stronger than 3.5 mJy (peak luminosity of $\sim$2$\times10^{10}$~Jy\,kpc$^2$) toward Arp 220, means that at present the known extragalactic methanol sources are not readily detectable at cosmologically interesting distances with current generation facilities. Even considering only the sources that meet our detection threshold in Figure \ref{fig:sensitivity}, we have four sources with star-formation rates (SFR) at the same level or higher than those observed toward NGC~253 or NGC~4945 with confident non-detections for 36.2-GHz class~I methanol masers.

When considering known examples of extragalactic class~I methanol masers, we see that locally, there is no clear link to star-formation rates in the regions where they are detected. The M1 region toward NGC~4945 and the masers toward NGC~253 are all likely associated with giant molecular clouds in these sources, however, these regions do not necessarily display enhanced star-formation rates when compared to the rest of the nuclear region \citep{Ellingsen+17b, McCarthy+18c}. M1 in NGC~4945 may be associated with the star-formation from Knot~B, though even if this is correct, we do not observe class~I maser emission from other regions in NGC~4945 with enhanced star-formation levels. Instead of tracing current star-formation, based on our current understanding of the environments of these masers (dense-gas and low-velocity shocks), we believe they may indicate regions that are soon to experience higher levels of star-formation, essentially the class~I maser emission is associated with the triggering of the star-formation.

The presence, or lack-thereof, of class~I methanol masers may not directly infer information about the star-formation rate of its host, however, it can still provide useful information about the mechanism triggering the star-formation. The morphological features that these class~I masers have been observed toward \citep{Ellingsen+17b, Gorski+18, McCarthy+18c} can be related to the star-formation in a galaxy and, therefore, the class~I masers (and their association with low-velocity shocks) can help identify mechanisms that are providing the gas fueling these starburst regions \citep{Ellingsen+17b}.


\section{Conclusions}

We present the results of new observations toward NGC~4945 and NGC~253, detecting new components of methanol emission toward both galaxies. Evidence of variability, especially toward the new region in NGC~4945, indicates that this emission is likely the result of a maser process. This is the first reported example of variability in an extragalactic class~I methanol maser and we recommend continued monitoring of these sources in order to both confirm and further quantify this variability.

We provide flux density upper-limits for the 36.2-GHz class~I and 37.7-GHz class~II methanol transitions toward 9 galaxies, observed as part of the original extragalactic class~I methanol searches which detected NGC~253 and NGC~4945, and 3 more sources observed more recently. In the majority of sources from the older observations (C2879), sensitivity levels are not sufficient in order to rule out maser emission of the same order as is observed toward NGC~4945 or NGC~253.

We also report the results of a re-analysis of the ATCA data on Arp~220. We do not find any evidence of the 36.2-GHz methanol maser emission toward this source. It appears a combination of poor data quality and sub-optimal imaging procedures may have resulted in the original reported detection. We place a $5\sigma$ upper limit on the 36.2-GHz line of 3.5~mJy.


\section*{Acknowledgements}

We thank the referee for the valuable comments which have helped improve the manuscript. The ATCA is part of the Australia Telescope which is funded by the Commonwealth of Australia for operation as a National Facility managed by CSIRO.  This research has made use of NASA's Astrophysics Data System Abstract Service. This research has made use of the NASA/IPAC Extragalactic Database (NED), which is operated by the Jet Propulsion Laboratory, California Institute of Technology, under contract with the National Aeronautics and Space Administration. This research also utilised {\sc aplpy}, an open-source plotting package for {\sc python} hosted at http://aplpy.github.com. This research made use of Astropy, a community-developed core Python package for Astronomy \citep{astropy+13}.

\bibliography{references}

\newcommand{\noop}[1]{}
\begin{thebibliography}{}
\makeatletter
\relax
\def\mn@urlcharsother{\let\do\@makeother \do\$\do\&\do\#\do\^\do\_\do\%\do\~}
\def\mn@doi{\begingroup\mn@urlcharsother \@ifnextchar [ {\mn@doi@}
  {\mn@doi@[]}}
\def\mn@doi@[#1]#2{\def\@tempa{#1}\ifx\@tempa\@empty \href
  {http://dx.doi.org/#2} {doi:#2}\else \href {http://dx.doi.org/#2} {#1}\fi
  \endgroup}
\def\mn@eprint#1#2{\mn@eprint@#1:#2::\@nil}
\def\mn@eprint@arXiv#1{\href {http://arxiv.org/abs/#1} {{\tt arXiv:#1}}}
\def\mn@eprint@dblp#1{\href {http://dblp.uni-trier.de/rec/bibtex/#1.xml}
  {dblp:#1}}
\def\mn@eprint@#1:#2:#3:#4\@nil{\def\@tempa {#1}\def\@tempb {#2}\def\@tempc
  {#3}\ifx \@tempc \@empty \let \@tempc \@tempb \let \@tempb \@tempa \fi \ifx
  \@tempb \@empty \def\@tempb {arXiv}\fi \@ifundefined
  {mn@eprint@\@tempb}{\@tempb:\@tempc}{\expandafter \expandafter \csname
  mn@eprint@\@tempb\endcsname \expandafter{\@tempc}}}

\bibitem[\protect\citeauthoryear{{Araya}, {Hofner}, {Goss}, {Kurtz},
  {Richards}, {Linz}, {Olmi}  \& {Sewi{\l}o}}{{Araya} et~al.}{2010}]{Araya+10}
{Araya} E.~D.,  {Hofner} P.,  {Goss} W.~M.,  {Kurtz} S.,  {Richards} A.~M.~S.,
  {Linz} H.,  {Olmi} L.,   {Sewi{\l}o} M.,  2010, \mn@doi [\apjl]
  {10.1088/2041-8205/717/2/L133}, \href
  {https://ui.adsabs.harvard.edu/abs/2010ApJ...717L.133A} {717, L133}

\bibitem[\protect\citeauthoryear{{Astropy Collaboration} et~al.,}{{Astropy
  Collaboration} et~al.}{2013}]{astropy+13}
{Astropy Collaboration} et~al., 2013, \mn@doi [\aap]
  {10.1051/0004-6361/201322068}, \href
  {http://adsabs.harvard.edu/abs/2013A%26A...558A..33A} {558, A33}

\bibitem[\protect\citeauthoryear{{Bendo}, {Henkel}, {D'Cruze}, {Dickinson},
  {Fuller}  \& {Karim}}{{Bendo} et~al.}{2016}]{Bendo+16}
{Bendo} G.~J.,  {Henkel} C.,  {D'Cruze} M.~J.,  {Dickinson} C.,  {Fuller}
  G.~A.,   {Karim} A.,  2016, \mn@doi [\mnras] {10.1093/mnras/stw1659}, \href
  {http://adsabs.harvard.edu/abs/2016MNRAS.463..252B} {463, 252}

\bibitem[\protect\citeauthoryear{{Bottinelli}, {Gouguenheim}, {Paturel}  \& {de
  Vaucouleurs}}{{Bottinelli} et~al.}{1985}]{Bottinelli+85}
{Bottinelli} L.,  {Gouguenheim} L.,  {Paturel} G.,   {de Vaucouleurs} G.,
  1985, \aaps, \href {https://ui.adsabs.harvard.edu/abs/1985A&AS...59...43B}
  {59, 43}

\bibitem[\protect\citeauthoryear{{Breen}, {Ellingsen}, {Caswell}  \&
  {Lewis}}{{Breen} et~al.}{2010}]{Breen+10a}
{Breen} S.~L.,  {Ellingsen} S.~P.,  {Caswell} J.~L.,   {Lewis} B.~E.,  2010,
  \mn@doi [\mnras] {10.1111/j.1365-2966.2009.15831.x}, \href
  {http://adsabs.harvard.edu/abs/2010MNRAS.401.2219B} {401, 2219}

\bibitem[\protect\citeauthoryear{{Breen}, {Ellingsen}, {Caswell}, {Green},
  {Fuller}, {Voronkov}, {Quinn}  \& {Avison}}{{Breen} et~al.}{2011}]{Breen+11}
{Breen} S.~L.,  {Ellingsen} S.~P.,  {Caswell} J.~L.,  {Green} J.~A.,  {Fuller}
  G.~A.,  {Voronkov} M.~A.,  {Quinn} L.~J.,   {Avison} A.,  2011, \mn@doi
  [\apj] {10.1088/0004-637X/733/2/80}, \href
  {http://adsabs.harvard.edu/abs/2011ApJ...733...80B} {733, 80}

\bibitem[\protect\citeauthoryear{{Breen}, {Ellingsen}, {Contreras}, {Green},
  {Caswell}, {Stevens}, {Dawson}  \& {Voronkov}}{{Breen}
  et~al.}{2013}]{Breen+13b}
{Breen} S.~L.,  {Ellingsen} S.~P.,  {Contreras} Y.,  {Green} J.~A.,  {Caswell}
  J.~L.,  {Stevens} J.~B.,  {Dawson} J.~R.,   {Voronkov} M.~A.,  2013, \mn@doi
  [\mnras] {10.1093/mnras/stt1315}, \href
  {http://adsabs.harvard.edu/abs/2013MNRAS.435..524B} {435, 524}

\bibitem[\protect\citeauthoryear{{Caswell}}{{Caswell}}{1997}]{Caswell97}
{Caswell} J.~L.,  1997, \mnras, \href
  {http://adsabs.harvard.edu/abs/1997MNRAS.289..203C} {289, 203}

\bibitem[\protect\citeauthoryear{{Chen}, {Ellingsen}, {Baan}, {Qiao}, {Li},
  {An}  \& {Breen}}{{Chen} et~al.}{2015}]{Chen+15}
{Chen} X.,  {Ellingsen} S.~P.,  {Baan} W.~A.,  {Qiao} H.-H.,  {Li} J.,  {An}
  T.,   {Breen} S.~L.,  2015, \mn@doi [\apjl] {10.1088/2041-8205/800/1/L2},
  \href {http://adsabs.harvard.edu/abs/2015ApJ...800L...2C} {800, L2}

\bibitem[\protect\citeauthoryear{{Chen}, {Ellingsen}, {Zhang}, {Wang}, {Shen},
  {Wu}  \& {Wu}}{{Chen} et~al.}{2016}]{Chen+16}
{Chen} X.,  {Ellingsen} S.~P.,  {Zhang} J.-S.,  {Wang} J.-Z.,  {Shen} Z.-Q.,
  {Wu} Q.-W.,   {Wu} Z.-Z.,  2016, \mn@doi [\mnras] {10.1093/mnras/stw680},
  \href {http://adsabs.harvard.edu/abs/2016MNRAS.459..357C} {459, 357}

\bibitem[\protect\citeauthoryear{{Chen}, {Ellingsen}, {Shen}, {McCarthy},
  {Zhong}  \& {Deng}}{{Chen} et~al.}{2018}]{Chen+18}
{Chen} X.,  {Ellingsen} S.~P.,  {Shen} Z.-Q.,  {McCarthy} T.~P.,  {Zhong}
  W.-Y.,   {Deng} H.,  2018, \mn@doi [\apjl] {10.3847/2041-8213/aab894}, \href
  {http://adsabs.harvard.edu/abs/2018ApJ...856L..35C} {856, L35}

\bibitem[\protect\citeauthoryear{{Chou} et~al.,}{{Chou} et~al.}{2007}]{Chou+07}
{Chou} R.~C.~Y.,  et~al., 2007, \mn@doi [\apj] {10.1086/521351}, \href
  {http://adsabs.harvard.edu/abs/2007ApJ...670..116C} {670, 116}

\bibitem[\protect\citeauthoryear{{Cyganowski}, {Brogan}, {Hunter}  \&
  {Churchwell}}{{Cyganowski} et~al.}{2009}]{Cyganowski+09}
{Cyganowski} C.~J.,  {Brogan} C.~L.,  {Hunter} T.~R.,   {Churchwell} E.,  2009,
  \mn@doi [\apj] {10.1088/0004-637X/702/2/1615}, \href
  {http://adsabs.harvard.edu/abs/2009ApJ...702.1615C} {702, 1615}

\bibitem[\protect\citeauthoryear{{Cyganowski}, {Brogan}, {Hunter}, {Zhang},
  {Friesen}, {Indebetouw}  \& {Chandler}}{{Cyganowski}
  et~al.}{2012}]{Cyganowski+12}
{Cyganowski} C.~J.,  {Brogan} C.~L.,  {Hunter} T.~R.,  {Zhang} Q.,  {Friesen}
  R.~K.,  {Indebetouw} R.,   {Chandler} C.~J.,  2012, preprint, \href
  {http://adsabs.harvard.edu/abs/2012arXiv1210.3366C} {} (\mn@eprint {arXiv}
  {1210.3366})

\bibitem[\protect\citeauthoryear{{Dalcanton} et~al.,}{{Dalcanton}
  et~al.}{2009}]{Dalcanton+09}
{Dalcanton} J.~J.,  et~al., 2009, \mn@doi [\apjs] {10.1088/0067-0049/183/1/67},
  \href {http://adsabs.harvard.edu/abs/2009ApJS..183...67D} {183, 67}

\bibitem[\protect\citeauthoryear{{Ellingsen}, {Breen}, {Caswell}, {Quinn}  \&
  {Fuller}}{{Ellingsen} et~al.}{2010}]{Ellingsen+10}
{Ellingsen} S.~P.,  {Breen} S.~L.,  {Caswell} J.~L.,  {Quinn} L.~J.,   {Fuller}
  G.~A.,  2010, \mn@doi [\mnras] {10.1111/j.1365-2966.2010.16349.x}, \href
  {http://adsabs.harvard.edu/abs/2010MNRAS.404..779E} {404, 779}

\bibitem[\protect\citeauthoryear{{Ellingsen}, {Chen}, {Qiao}, {Baan}, {An},
  {Li}  \& {Breen}}{{Ellingsen} et~al.}{2014}]{Ellingsen+14}
{Ellingsen} S.~P.,  {Chen} X.,  {Qiao} H.-H.,  {Baan} W.,  {An} T.,  {Li} J.,
  {Breen} S.~L.,  2014, \mn@doi [\apjl] {10.1088/2041-8205/790/2/L28}, \href
  {http://adsabs.harvard.edu/abs/2014ApJ...790L..28E} {790, L28}

\bibitem[\protect\citeauthoryear{{Ellingsen}, {Chen}, {Breen}  \&
  {Qiao}}{{Ellingsen} et~al.}{2017}]{Ellingsen+17b}
{Ellingsen} S.~P.,  {Chen} X.,  {Breen} S.~L.,   {Qiao} H.-H.,  2017, \mn@doi
  [\mnras] {10.1093/mnras/stx2076}, \href
  {http://adsabs.harvard.edu/abs/2017MNRAS.472..604E} {472, 604}

\bibitem[\protect\citeauthoryear{Felli et~al.,}{Felli et~al.}{2007}]{Felli+07}
Felli M.,  et~al., 2007, \mn@doi [Astronomy & Astrophysics]
  {10.1051/0004-6361:20077804}, 476, 373–664

\bibitem[\protect\citeauthoryear{{Gao} et~al.,}{{Gao} et~al.}{2016}]{Gao+16}
{Gao} F.,  et~al., 2016, \mn@doi [\apj] {10.3847/0004-637X/817/2/128}, \href
  {https://ui.adsabs.harvard.edu/abs/2016ApJ...817..128G} {817, 128}

\bibitem[\protect\citeauthoryear{{Goedhart}, {Gaylard}  \& {van der
  Walt}}{{Goedhart} et~al.}{2003}]{Goedhart+03}
{Goedhart} S.,  {Gaylard} M.~J.,   {van der Walt} D.~J.,  2003, \mn@doi
  [\mnras] {10.1046/j.1365-8711.2003.06426.x}, \href
  {https://ui.adsabs.harvard.edu/abs/2003MNRAS.339L..33G} {339, L33}

\bibitem[\protect\citeauthoryear{{Gorski}, {Ott}, {Rand}, {Meier}, {Momjian}
  \& {Schinnerer}}{{Gorski} et~al.}{2017}]{Gorski+17}
{Gorski} M.,  {Ott} J.,  {Rand} R.,  {Meier} D.~S.,  {Momjian} E.,
  {Schinnerer} E.,  2017, \mn@doi [\apj] {10.3847/1538-4357/aa74af}, \href
  {http://adsabs.harvard.edu/abs/2017ApJ...842..124G} {842, 124}

\bibitem[\protect\citeauthoryear{{Gorski}, {Ott}, {Rand}, {Meier}, {Momjian}
  \& {Schinnerer}}{{Gorski} et~al.}{2018}]{Gorski+18}
{Gorski} M.,  {Ott} J.,  {Rand} R.,  {Meier} D.~S.,  {Momjian} E.,
  {Schinnerer} E.,  2018, \mn@doi [\apj] {10.3847/1538-4357/aab3cc}, \href
  {http://adsabs.harvard.edu/abs/2018ApJ...856..134G} {856, 134}

\bibitem[\protect\citeauthoryear{{Green} et~al.,}{{Green}
  et~al.}{2008}]{Green+08}
{Green} J.~A.,  et~al., 2008, \mn@doi [\mnras]
  {10.1111/j.1365-2966.2008.12888.x}, \href
  {http://adsabs.harvard.edu/abs/2008MNRAS.385..948G} {385, 948}

\bibitem[\protect\citeauthoryear{{Haschick} \& {Baan}}{{Haschick} \&
  {Baan}}{1990}]{Haschick+90}
{Haschick} A.~D.,  {Baan} W.~A.,  1990, \mn@doi [\apjl] {10.1086/185729}, \href
  {https://ui.adsabs.harvard.edu/abs/1990ApJ...355L..23H} {355, L23}

\bibitem[\protect\citeauthoryear{{Henkel} et~al.,}{{Henkel}
  et~al.}{2018}]{Henkel+18}
{Henkel} C.,  et~al., 2018, \mn@doi [\aap] {10.1051/0004-6361/201732174}, \href
  {http://adsabs.harvard.edu/abs/2018A%26A...615A.155H} {615, A155}

\bibitem[\protect\citeauthoryear{{Hollenbach}, {Elitzur}  \&
  {McKee}}{{Hollenbach} et~al.}{2013}]{Hollenbach+13}
{Hollenbach} D.,  {Elitzur} M.,   {McKee} C.~F.,  2013, preprint, \href
  {http://adsabs.harvard.edu/abs/2013arXiv1306.5276H} {} (\mn@eprint {arXiv}
  {1306.5276})

\bibitem[\protect\citeauthoryear{{Jensen}, {Tonry}, {Barris}, {Thompson},
  {Liu}, {Rieke}, {Ajhar}  \& {Blakeslee}}{{Jensen} et~al.}{2003}]{Jensen+03}
{Jensen} J.~B.,  {Tonry} J.~L.,  {Barris} B.~J.,  {Thompson} R.~I.,  {Liu}
  M.~C.,  {Rieke} M.~J.,  {Ajhar} E.~A.,   {Blakeslee} J.~P.,  2003, \mn@doi
  [\apj] {10.1086/345430}, \href
  {https://ui.adsabs.harvard.edu/abs/2003ApJ...583..712J} {583, 712}

\bibitem[\protect\citeauthoryear{{Kurtz}, {Hofner}  \& {{\'A}lvarez}}{{Kurtz}
  et~al.}{2004}]{Kurtz+04}
{Kurtz} S.,  {Hofner} P.,   {{\'A}lvarez} C.~V.,  2004, \mn@doi [\apjs]
  {10.1086/423956}, \href {http://adsabs.harvard.edu/abs/2004ApJS..155..149K}
  {155, 149}

\bibitem[\protect\citeauthoryear{{Leurini}, {Menten}  \& {Walmsley}}{{Leurini}
  et~al.}{2016}]{Leurini+16}
{Leurini} S.,  {Menten} K.~M.,   {Walmsley} C.~M.,  2016, \mn@doi [\aap]
  {10.1051/0004-6361/201527974}, \href
  {http://adsabs.harvard.edu/abs/2016A%26A...592A..31L} {592, A31}

\bibitem[\protect\citeauthoryear{{McCallum}, {Ellingsen}  \&
  {Lovell}}{{McCallum} et~al.}{2007}]{McCallum+07}
{McCallum} J.~N.,  {Ellingsen} S.~P.,   {Lovell} J.~E.~J.,  2007, \mn@doi
  [\mnras] {10.1111/j.1365-2966.2007.11492.x}, \href
  {https://ui.adsabs.harvard.edu/abs/2007MNRAS.376..549M} {376, 549}

\bibitem[\protect\citeauthoryear{{McCarthy}, {Ellingsen}, {Chen}, {Breen},
  {Voronkov}  \& {Qiao}}{{McCarthy} et~al.}{2017}]{McCarthy+17}
{McCarthy} T.~P.,  {Ellingsen} S.~P.,  {Chen} X.,  {Breen} S.~L.,  {Voronkov}
  M.~A.,   {Qiao} H.-h.,  2017, \mn@doi [\apj] {10.3847/1538-4357/aa872c},
  \href {http://adsabs.harvard.edu/abs/2017ApJ...846..156M} {846, 156}

\bibitem[\protect\citeauthoryear{{McCarthy}, {Ellingsen}, {Breen}, {Henkel},
  {Voronkov}  \& {Chen}}{{McCarthy} et~al.}{2018a}]{McCarthy+18c}
{McCarthy} T.~P.,  {Ellingsen} S.~P.,  {Breen} S.~L.,  {Henkel} C.,  {Voronkov}
  M.~A.,   {Chen} X.,  2018a, \mn@doi [\mnras] {10.1093/mnras/sty2192}, \href
  {http://adsabs.harvard.edu/abs/2018MNRAS.480.4578M} {480, 4578}

\bibitem[\protect\citeauthoryear{{McCarthy}, {Ellingsen}, {Breen}, {Voronkov}
  \& {Chen}}{{McCarthy} et~al.}{2018b}]{McCarthy+18b}
{McCarthy} T.~P.,  {Ellingsen} S.~P.,  {Breen} S.~L.,  {Voronkov} M.~A.,
  {Chen} X.,  2018b, \mn@doi [\apjl] {10.3847/2041-8213/aae82c}, \href
  {http://adsabs.harvard.edu/abs/2018ApJ...867L...4M} {867, L4}

\bibitem[\protect\citeauthoryear{{McEwen}, {Pihlstr{\"o}m}  \&
  {Sjouwerman}}{{McEwen} et~al.}{2014}]{McEwen+14}
{McEwen} B.~C.,  {Pihlstr{\"o}m} Y.~M.,   {Sjouwerman} L.~O.,  2014, \mn@doi
  [\apj] {10.1088/0004-637X/793/2/133}, \href
  {http://adsabs.harvard.edu/abs/2014ApJ...793..133M} {793, 133}

\bibitem[\protect\citeauthoryear{{Moran}, {Greenhill}  \& {Herrnstein}}{{Moran}
  et~al.}{1999}]{Moran+99}
{Moran} J.~M.,  {Greenhill} L.~J.,   {Herrnstein} J.~R.,  1999, \mn@doi
  [Journal of Astrophysics and Astronomy] {10.1007/BF02702350}, \href
  {http://adsabs.harvard.edu/abs/1999JApA...20..165M} {20, 165}

\bibitem[\protect\citeauthoryear{{Olech}, {Szymczak}, {Wolak}, {Sarniak}  \&
  {Bartkiewicz}}{{Olech} et~al.}{2019}]{Olech+19}
{Olech} M.,  {Szymczak} M.,  {Wolak} P.,  {Sarniak} R.,   {Bartkiewicz} A.,
  2019, \mn@doi [\mnras] {10.1093/mnras/stz926}, \href
  {https://ui.adsabs.harvard.edu/abs/2019MNRAS.486.1236O} {486, 1236}

\bibitem[\protect\citeauthoryear{{Sakamoto}, {Mao}, {Matsushita}, {Peck},
  {Sawada}  \& {Wiedner}}{{Sakamoto} et~al.}{2011}]{Sakamoto+11}
{Sakamoto} K.,  {Mao} R.-Q.,  {Matsushita} S.,  {Peck} A.~B.,  {Sawada} T.,
  {Wiedner} M.~C.,  2011, \mn@doi [\apj] {10.1088/0004-637X/735/1/19}, \href
  {http://adsabs.harvard.edu/abs/2011ApJ...735...19S} {735, 19}

\bibitem[\protect\citeauthoryear{{Sjouwerman}, {Murray}, {Pihlstr{\"o}m},
  {Fish}  \& {Araya}}{{Sjouwerman} et~al.}{2010}]{Sjouwerman+10}
{Sjouwerman} L.~O.,  {Murray} C.~E.,  {Pihlstr{\"o}m} Y.~M.,  {Fish} V.~L.,
  {Araya} E.~D.,  2010, \mn@doi [\apjl] {10.1088/2041-8205/724/2/L158}, \href
  {http://adsabs.harvard.edu/abs/2010ApJ...724L.158S} {724, L158}

\bibitem[\protect\citeauthoryear{{Sobolev} et~al.,}{{Sobolev}
  et~al.}{2007}]{Sobolev+07}
{Sobolev} A.~M.,  et~al., 2007, in {Chapman} J.~M.,  {Baan} W.~A.,  eds,  IAU
  Symposium Vol. 242, Astrophysical Masers and their Environments. pp 81--88
  (\mn@eprint {arXiv} {0706.3117}), \mn@doi{10.1017/S1743921307012616}

\bibitem[\protect\citeauthoryear{{Sorce}, {Tully}, {Courtois}, {Jarrett},
  {Neill}  \& {Shaya}}{{Sorce} et~al.}{2014}]{Sorce+14}
{Sorce} J.~G.,  {Tully} R.~B.,  {Courtois} H.~M.,  {Jarrett} T.~H.,  {Neill}
  J.~D.,   {Shaya} E.~J.,  2014, \mn@doi [\mnras] {10.1093/mnras/stu1450},
  \href {https://ui.adsabs.harvard.edu/abs/2014MNRAS.444..527S} {444, 527}

\bibitem[\protect\citeauthoryear{{Tsunekawa}, {Ukai}, {Toyama}  \&
  {Takagi}}{{Tsunekawa} et~al.}{1995}]{Tsunekawa+95}
{Tsunekawa} S.,  {Ukai} T.,  {Toyama} A.,   {Takagi} K.,  1995, Department of
  Physics, Toyama Univerity, Japan

\bibitem[\protect\citeauthoryear{{Tully} \& {Fisher}}{{Tully} \&
  {Fisher}}{1988}]{Tully+88}
{Tully} R.~B.,  {Fisher} J.~R.,  1988, {Catalog of Nearby Galaxies}

\bibitem[\protect\citeauthoryear{{Tully}, {Rizzi}, {Shaya}, {Courtois},
  {Makarov}  \& {Jacobs}}{{Tully} et~al.}{2009}]{Tully+09}
{Tully} R.~B.,  {Rizzi} L.,  {Shaya} E.~J.,  {Courtois} H.~M.,  {Makarov}
  D.~I.,   {Jacobs} B.~A.,  2009, \mn@doi [\aj] {10.1088/0004-6256/138/2/323},
  \href {https://ui.adsabs.harvard.edu/abs/2009AJ....138..323T} {138, 323}

\bibitem[\protect\citeauthoryear{{Tully} et~al.,}{{Tully}
  et~al.}{2013}]{Tully+13}
{Tully} R.~B.,  et~al., 2013, \mn@doi [\aj] {10.1088/0004-6256/146/4/86}, \href
  {http://adsabs.harvard.edu/abs/2013AJ....146...86T} {146, 86}

\bibitem[\protect\citeauthoryear{{Tully}, {Courtois}  \& {Sorce}}{{Tully}
  et~al.}{2016}]{Tully+16}
{Tully} R.~B.,  {Courtois} H.~M.,   {Sorce} J.~G.,  2016, \mn@doi [\aj]
  {10.3847/0004-6256/152/2/50}, \href
  {https://ui.adsabs.harvard.edu/abs/2016AJ....152...50T} {152, 50}

\bibitem[\protect\citeauthoryear{{Voronkov}, {Brooks}, {Sobolev}, {Ellingsen},
  {Ostrovskii}  \& {Caswell}}{{Voronkov} et~al.}{2006}]{Voronkov+06}
{Voronkov} M.~A.,  {Brooks} K.~J.,  {Sobolev} A.~M.,  {Ellingsen} S.~P.,
  {Ostrovskii} A.~B.,   {Caswell} J.~L.,  2006, \mn@doi [\mnras]
  {10.1111/j.1365-2966.2006.11047.x}, \href
  {http://adsabs.harvard.edu/abs/2006MNRAS.373..411V} {373, 411}

\bibitem[\protect\citeauthoryear{{Voronkov}, {Caswell}, {Ellingsen}, {Green}
  \& {Breen}}{{Voronkov} et~al.}{2014}]{Voronkov+14}
{Voronkov} M.~A.,  {Caswell} J.~L.,  {Ellingsen} S.~P.,  {Green} J.~A.,
  {Breen} S.~L.,  2014, \mn@doi [\mnras] {10.1093/mnras/stu116}, \href
  {http://adsabs.harvard.edu/abs/2014MNRAS.439.2584V} {439, 2584}

\bibitem[\protect\citeauthoryear{{Wang}, {Zhang}, {Gao}, {Zhang}, {Li}, {Fang}
  \& {Shi}}{{Wang} et~al.}{2014}]{Wang+14}
{Wang} J.,  {Zhang} J.,  {Gao} Y.,  {Zhang} Z.-Y.,  {Li} D.,  {Fang} M.,
  {Shi} Y.,  2014, \mn@doi [Nature Communications] {10.1038/ncomms6449}, \href
  {http://adsabs.harvard.edu/abs/2014NatCo...5E5449W} {5, 5449}

\bibitem[\protect\citeauthoryear{{Whiting}}{{Whiting}}{2012}]{Whiting12}
{Whiting} M.~T.,  2012, \mn@doi [\mnras] {10.1111/j.1365-2966.2012.20548.x},
  \href {https://ui.adsabs.harvard.edu/abs/2012MNRAS.421.3242W} {421, 3242}

\bibitem[\protect\citeauthoryear{{Wilson} et~al.,}{{Wilson}
  et~al.}{2011}]{Wilson+11}
{Wilson} W.~E.,  et~al., 2011, \mn@doi [\mnras]
  {10.1111/j.1365-2966.2011.19054.x}, \href
  {http://adsabs.harvard.edu/abs/2011MNRAS.416..832W} {416, 832}

\bibitem[\protect\citeauthoryear{{Yamauchi}, {Nakai}, {Ishihara}, {Diamond}  \&
  {Sato}}{{Yamauchi} et~al.}{2012}]{Yamauchi+12}
{Yamauchi} A.,  {Nakai} N.,  {Ishihara} Y.,  {Diamond} P.,   {Sato} N.,  2012,
  \mn@doi [\pasj] {10.1093/pasj/64.5.103}, \href
  {https://ui.adsabs.harvard.edu/abs/2012PASJ...64..103Y} {64, 103}

\bibitem[\protect\citeauthoryear{{Yusef-Zadeh}, {Cotton}, {Viti}, {Wardle}  \&
  {Royster}}{{Yusef-Zadeh} et~al.}{2013}]{Yusef-Zadeh+13}
{Yusef-Zadeh} F.,  {Cotton} W.,  {Viti} S.,  {Wardle} M.,   {Royster} M.,
  2013, \mn@doi [\apjl] {10.1088/2041-8205/764/2/L19}, \href
  {http://adsabs.harvard.edu/abs/2013ApJ...764L..19Y} {764, L19}

\bibitem[\protect\citeauthoryear{{van der Walt}}{{van der
  Walt}}{2011}]{vanderwalt11}
{van der Walt} D.~J.,  2011, \mn@doi [\aj] {10.1088/0004-6256/141/5/152}, \href
  {https://ui.adsabs.harvard.edu/abs/2011AJ....141..152V} {141, 152}

\makeatother
\end{thebibliography}

\onecolumn

\appendix

\setcounter{table}{0} 
\renewcommand{\thetable}{A.\arabic{table}} 

\begin{table*} 
	\begin{center}
		\caption{Details of the 2014 March sources excluded due to poor data quality. Recession velocities have been taken from the \textit{NASA/IPAC Extragalactic Database}$^\dagger$~(NED). Redshift-independent distances have been provided where possible, otherwise distances reported are the luminosity distances provided by NED. }
		\begin{tabular}{@{}lccll@{}}
			\toprule
			\multicolumn{1}{c}{Target}  & \multicolumn{1}{c}{Right Ascension}  &\multicolumn{1}{c}{Declination}     & \multicolumn{1}{c}{$V_{helio}$} & \multicolumn{1}{c}{$D$}  \\
			\multicolumn{1}{c}{Source} &   \multicolumn{1}{c}{\bf $h$~~~$m$~~~$s$}& \multicolumn{1}{c}{\bf $^\circ$~~~$\prime$~~~$\prime\prime$} &\multicolumn{1}{c}{(\kms)}  &\multicolumn{1}{c}{(Mpc)}   \\ \midrule
			MRK\,1029          & 02 17 03.57 &  +05 17 31.40 & 9076 & 133  \\
			NGC\,1052              & 02 41 04.80  & --08 15 21.00 & 1510 & 18.0$^{[1]}$   \\
			J0350--0127                 & 03 50 00.35 & --01 27 57.70 & 12322 & 185  \\                   
			UGCA\,116             & 05 55 42.63 &  +03 23 31.80 & 789 & 10.3$^{[2]}$  \\
			NGC\,5765b              & 14 50 51.50 & +05 06 52.00 & 8333 & 126$^{[3]}$   \\
			TXS\,2226--184             & 22 29 12.50 & --18 10 47.00 & 7520 & 108  \\  \hline
		\end{tabular} \label{tab:excluded_sources}
	\end{center}
\begin{flushleft}
	Note: $^{[1]}$\citet{Jensen+03}, $^{[2]}$\citet{Tully+88}, and $^{[3]}$\citet{Gao+16}  \\
\end{flushleft}	

\end{table*}

\end{document}